\documentclass[preprints,article,submit,moreauthors,pdftex]{mdpi}

\pdfoutput=1 
\usepackage{color}
\usepackage{graphicx}
\usepackage{dcolumn}
\usepackage{bm}
\usepackage{braket}
\firstpage{1} 
\pubvolume{1}
\issuenum{1}
\articlenumber{0}
\pubyear{2021}
\copyrightyear{2021}
\datereceived{} 
\dateaccepted{} 
\datepublished{} 
\hreflink{https://doi.org/} % If needed use \linebreak
\Title{Majorana Zero Modes in Ferromagnetic Wires without Spin-Orbit Coupling}

\TitleCitation{Majorana Zero Modes in Ferromagnetic Wires without Spin-Orbit Coupling}

\newcommand{\be}[1]{\begin{equation}\label{eq:#1}}
\newcommand{\ee}{\end{equation}}
\newcommand{\bea}{\begin{eqnarray}}
\newcommand{\eea}{\end{eqnarray}}

\newcommand{\phd}{\phantom{\dag}}

\newcommand{\up}{^{\phd}}
\newcommand{\noi}{\noindent}
\newcommand{\no}{\nonumber}

\Author{Giorgos Livanas %MDPI: Please carefully check the accuracy of names and affiliations. 
, Nikolaos Vanas and Georgios Varelogiannis}

\AuthorNames{Giorgos Livanas, Nikolaos Vanas and Georgios Varelogiannis}

\AuthorCitation{Livanas, G.; Vanas, N.; Varelogiannis, G.}
\address[1]{%
Department of Applied Mathematical and Physical Sciences, National Technical University of Athens,\linebreak 15780 Athens, Greece; n.vanas@protonmail.com (N.V.); varelogi@mail.ntua.gr (G.V.)}

\corres{\hangafter=1 \hangindent=1.05em \hspace{-0.82em}Correspondence: livanasg@mail.ntua.gr;}

\abstract{We present a novel controllable platform for engineering Majorana zero modes. The platform consists of a ferromagnetic metallic wire placed among conventional superconductors, which are in proximity to ferromagnetic insulators. We demonstrate that Majorana zero modes emerge localised at the edges of the ferromagnetic wire, due to the interplay of the applied supercurrents and the induced by proximity exchange fields with conventional superconductivity.  Our mechanism does not rely on the pairing of helical fermions by combining conventional superconductivity with spin-orbit coupling, but rather exploits the misalignment between the magnetization of the ferromagnetic insulators and that of the ferromagnetic wire.}


\begin{document}
%%%%%%%%%%%%%%%%%%%%%%%%%%%%%%%%%%%%%%%%%%
%\setcounter{section}{-1} %% Remove this when starting to work on the template.
%\section{How to Use this Template}

\section{Introduction}

The development of quantum computers promises a new technological revolution \cite{ref-journal1}. However, a fundamental obstacle hindering the development of quantum computation is quantum decoherence, the loss of quantum mechanical phase coherence and, therefore, of information encoded in qubits. \textls[-10]{Attempts to deal with this problem through quantum error correction algorithms lead to elaborate schemes consuming a highly disproportional amount of qubits. On the contrary, topological quantum computers suppress quantum decoherence at the hardware level. Qubits in these devices are realized from topologically protected entities, which are immune to environmental noise and, therefore, decoherence~\cite{ref-journal2,ref-journal3,ref-journal4}}. 

Such topological protected entities are Majorana zero modes---massless neutral particles that constitute their own antiparticles. Characterised by their particle--antiparticle symmetry, MZMs emerge as quasiparticles bound to defects or boundaries of topological superconductors \cite{ref-journal5}.
The simplest topological superconductor is the effectively spinless superconducting phase of \emph{p}-wave symmetry \cite{ref-journal6,ref-journal7}. Although a topological $p$-wave state is proposed for the superconducting phase of Sr$_2$RuO$_4$ \cite{ref-journal8}, experimental results are yet inconclusive \cite{ref-journal9}. 

Instead, several proposals have been put forward for engineering topological \emph{p}-wave superconductivity out of more trivial materials. Among them, we distinguish conventional superconductor/topological insulator heterostructures \cite{ref-journal10}, semiconductors in proximity to conventional superconductors \cite{ref-journal11}, non-centrosymmetric superconductors \cite{ref-journal12} and planar Josephson junctions \cite{ref-journal13}.  All these proposals effectuate the pairing of helical fermions by combining conventional superconductivity with spin orbit coupling in the presence of a spin-splitting field, in order to realize an effectively spinless superconducting state.

%%%%%%%%%%%%%%%%%%%%%%%%%%%%%%%%%%%%%%%%%%
\section{Majorana Zero Modes in Superconductor/Ferromagnet Heterostructures}

As aforementioned, realising MZMs using conventional superconductors requires the presence of a spin-splitting field. The simplest way to satisfy this requirement is by applying an external magnetic field. However, magnetic fields are detrimental to conventional superconductivity; thereby, many proposals for realising MZMs are based on heterostructures among superconductors with ferromagnetic materials. Superconductor-ferromagnet heterostructures do not require the application of an external magnetic field and, therefore, are advantageous with respect to other platforms. \textls[+10]{Following this line of argument, topological superconductivity and MZMs have been demonstrated to emerge in half-metal/superconductor \cite{ref-journal14} or ferromagnet/unconventional superconductor \cite{ref-journal15} heterostructures, in ferromagnetic wires proximised to conventional \mbox{superconductors~\cite{ref-journal16,ref-journal17,ref-journal18}} and in ferromagnetically aligned chains of magnetic impurities embedded in conventional superconductors \cite{ref-journal19,ref-journal20,ref-journal21,ref-journal22,ref-journal23,ref-journal24,ref-journal25,ref-journal26}. Spin-orbit coupling is an essential component in all of these~proposals}. 

\subsection{A Novel Platform}

\begin{figure}[h]
\includegraphics[scale=0.6]{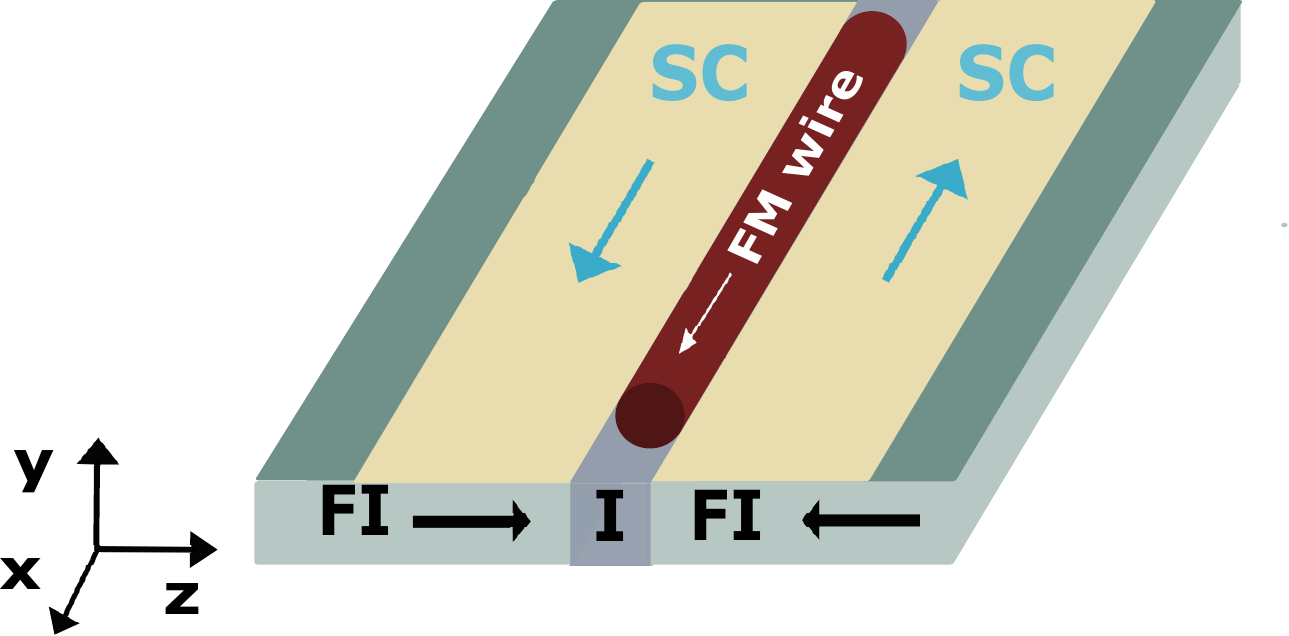}
\includegraphics[scale=0.2]{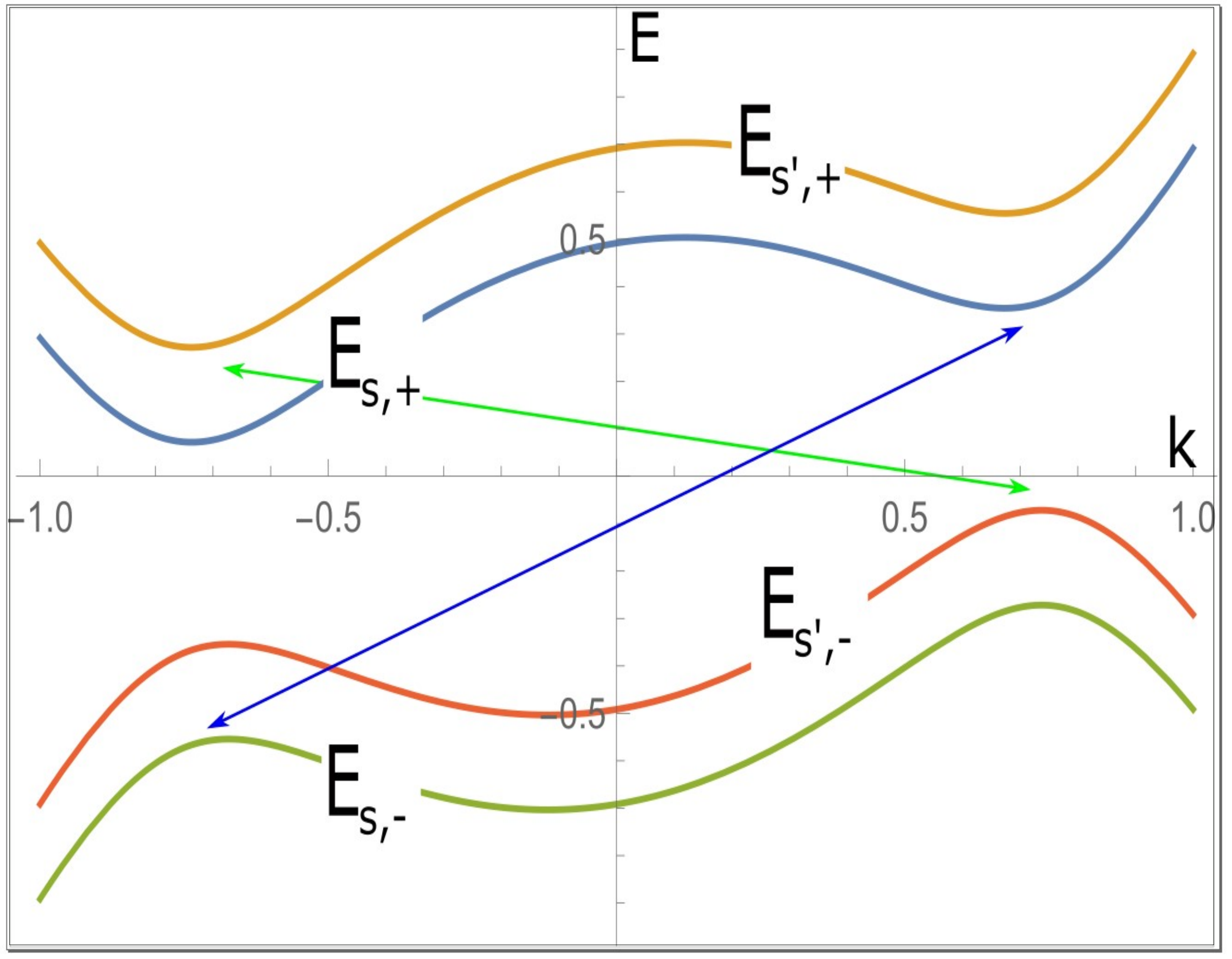}
\caption{ (\textbf{Left}) Novel platform for engineering Majorana zero modes. A ferromagnetic metallic wire placed among two conventional superconductors where supercurrents flow in opposite directions (blue arrows). The superconducting layers are deposited upon ferromagnetic insulators with opposite magnetisations (black arrows), which are perpendicular to the magnetisation of the ferromagnetic wire (white arrow).  (\textbf{Right}) The energy bands, $E_{s,\pm}$ where $s$ refers to spin components and $\pm$ to hole and particles, respectively, of a conventional superconductor when a supercurrent and an exchange field are applied. Supercurrent breaks inversion symmetry ($k$ , $-k$ momenta are not equivalent) and the exchange field breaks spin symmetry ($s$ and $s'$ are not equivalent)  leading to the emergence of  $p$-wave superconducting correlations. Arrows indicate the pairing components of conventional superconductivity. \label{fig1}}
\end{figure}

Here, we put forward a novel platform for engineering topological superconductivity and Majorana zero modes, which does not require spin orbit coupling. The platform, as presented in Figure \ref{fig1}, consists of a ferromagnetic metallic wire, which is placed between two conventional superconducting layers where supercurrents flow in opposite directions. The superconducting layers are deposited on  ferromagnetic insulators. The magnetisations of the two ferromagnetic insulators point in opposite directions and both are perpendicular to the magnetisation of the ferromagnetic metallic wire. 
Majorana zero modes emerge localised at the edges of the FM, due to the synergy between the externally applied supercurrents with the emergent in the SCs exchange fields~\cite{ref-journal27}. We remark that several works have demonstrated both theoretically \cite{ref-journal28,ref-journal29} and experimentally \cite{ref-journal30,ref-journal31,ref-journal32} the emergence of exchange fields in conventional SCs proximised by FIs. Moreover, developments in the fabrication of SC-FI heterostructures \cite{ref-journal33} and ferromagnetic metallic nanonwires embedded in conventional SCs \cite{ref-journal34,ref-journal35,ref-journal36}, enhance the experimental feasibility of the proposed platform.

\subsection{The Underlying Mechanism}

According to our mechanism, a topological superconducting state is stabilized over the FM wire, when charge supercurrents are applied in conventional SCs where finite exchange fields are present. The supercurrent and the exchange field break inversion and spin symmetries, respectively, and, therefore, partially convert conventional superconductivity to a $p$-wave superconducting field. This emergent $p$-wave superconducting field still pairs electrons of different energy bands of the superconductor, while MZMs require intraband superconductivity.  

However, in an FM wire with magnetisation perpendicular to the magnetisation of the ferromagnetic insulators, the induced $p$-wave superconducting field pairs electrons of the same energy band of the FM. An intraband superconducting component emerges when the $d$-vector of $p$-wave superconductivity \cite{ref-journal37} is misaligned to the exchange field of the FM wire \cite{ref-journal38}. Thus, when this $p$-wave superconducting field mediates into the FM wire, due to the proximity effect, an effectively spinless superconducting state is stabilised, and MZMs emerge. In the next section, we present analytically how the triplet superconducting component emerges, due to the coexistence of the exchange field and the supercurrents with  conventional superconductivity.

%%%%%%%%%%%%%%%%%%%%%%%%%%%%%%%%%%%%%%%%%%

%%%%%%%%%%%%%%%%%%%%%%%%%%%%%%%%%%%%%%%%%%
\section{Induced \emph{p}-Wave Superconductivity}

In order to demonstrate how triplet $p$-wave superconductivity is induced, we  consider a superconducting wire with order parameter $\Delta$, where a supercurrent $J$ is applied in the presence of an exchange field $h_z$. We  consider the following Hamiltonian 
\bea
 H=\int dx \Psi^{\dag}(x) \left( [\frac{\partial_x^2}{2m} \right. +\mu]\tau_z
 + h_z\tau_z\sigma_z +\Delta\tau_y\sigma_y  \biggl ) \Psi(x) \, \label{Eq:1}
\eea

\noi where $\Psi^{\dag}=(\psi_{\uparrow}^{\dag},\psi_{\downarrow}^{\dag},\psi_{\uparrow},\psi_{\downarrow})$ the extended Nambu spinor and $\bm{\tau}$ and $\bm{\sigma}$ are the Pauli matrices acting on particle-hole and spin space, respectively, { $\psi_{s}^{(\dag)}(x)$ the operator destroying(creating) an electron with spin s at coordinate x,} $h_z$ the exchange field and $\Delta=|\Delta|e^{iJx\tau_z}$ the singlet superconducting field.  Under the gauge transformation 
\bea 
\Psi(x) \rightarrow e^{-iJ/2x\tau_z}\Psi(x)
\eea

\noi  kinetic term $\frac{\partial_x^2}{2m}$ transforms to $\frac{\partial_x^2}{2m}+i\frac{J}{2m}\partial_x - \frac{J^2}{8m}\tau_z$ where $i\frac{J}{2m}\partial_x$ is a current term and $- \frac{J^2}{8m}\tau_z$ can be absorbed in the chemical potential $\mu$. Thus, Hamiltonian Equation (\ref{Eq:1}) takes the form 
\bea
 H=\int dx \Psi^{\dag}(x) \biggl( [\frac{\partial_x^2}{2m}-\mu]\tau_z  +i\frac{J}{2m}\partial_x + h_z\tau_z\sigma_z +\Delta \tau_y\sigma_y  \biggl ) \Psi(x) \label{Eq:2}
\eea

%\begin{center}
%\begin{figure}[H]
%\begin{tabular}{c}
%\includegraphics[scale=0.2]{figs/mhJDcontinuumb.eps}
%\end{tabular}
%\caption{The energy bands of a conventional superconductor where a supercurrent and an exchange field are applied.  \label{fig2}}
%\end{figure} 
%\end{center}

\noi {Considering a translationally} invariant wire Hamiltonian Equation (\ref{Eq:2}) takes the following form upon the Fourier transformation 
\bea
\Psi^{\dag}(x)=\int e^{-ikx}\Psi(k)^{\dag}dk
\eea

\noi to momenta $k$ space, where $\int \Psi^{\dag}(x)\Psi(x)=\int \Psi^{\dag}(k)\Psi(k)=1$.
\bea
 H=\int dk \Psi^{\dag}(k) \hat{H}(k) \Psi(k)=\int dk \Psi^{\dag}(k) \biggl( [\frac{k^2}{2m}-\mu]\tau_z  + \frac{J}{2m}k + h_z\tau_z\sigma_z +\Delta \tau_y\sigma_y  \biggl ) \Psi(k) \label{Eq:3}
\eea

\noi The energy bands of the wire 
\bea
E_{s,\pm}(k)= -sh_z + \frac{J}{2m}k \pm \sqrt {\left (\frac{k^2}{2m}-\mu \right )^2 + \Delta^2}
\eea

\noi where $s=\pm$ corresponds to $\downarrow$ and $\uparrow$ bands respectively, derive from the transformation 
\bea
U(k)=\frac{\sqrt{2}}{2}\biggl[ \left [A^{-}(k)(\tau_z+\tau_0)+A^{+}(k)(\tau_x+i\tau_y) \right ] e^{i\frac{\pi}{4}\sigma_y}+(\tau_x-i\tau_y)+ (\tau_0-\tau_z) \biggl ]
\eea

\noi and 
\vspace{6pt}
\bea
U(k)^{-1}=\frac{\sqrt{2}}{2}\biggl [A^{-}(k)(\tau_0-\tau_z)+A^{+}(k)(\tau_x-i\tau_y)+B(k)(\tau_x+i\tau_y + \tau_0+\tau_z)e^{-i\frac{\pi}{4}\sigma_y} \biggl]
\eea

\noi diagonalising Hamiltonian matrix $\hat{H}(k)$, where 
\bea
A^{\pm}(k)=\frac{\frac{k^2}{2m}-\mu \pm \sqrt{(\frac{k^2}{2m}-\mu)^2+\Delta^2}}{\Delta}
\eea

\noi and 
\bea
B(k)=\frac{\Delta}{\sqrt{(\frac{k^2}{2m}-\mu)^2+\Delta^2}}
\eea

\noi The induced \emph{p}-wave correlations derive from the equation 
\bea
<\Delta_{\bm{p}}> \propto \sum_{k}\sum_{m} n_{F}(E_m)[U(k)^{\dag} k\tau_x \bm{d} \cdot \tilde{\bm{\sigma}}i\sigma_y U(k)]_{mm}
\eea

\noi { where $n_{F}(E_m)$ the Fermi distribution and $\tilde{\bm{\sigma}}=(\sigma_x,\tau_z\sigma_y,\sigma_z)$}. For $\bm{d}=d_z=(0,0,1)$, we find 
\bea
<\Delta_{p_z}> \propto \sum_{k}\sum_{m} k n_{F}(E_m)[U(k)^{\dag}\tau_x\sigma_x U(k)]_{mm}
\eea

\noi where 
\bea
[U(k)^{\dag}\tau_x\sigma_x U(k)]= \frac{\Delta}{\sqrt{(\frac{k^2}{2m}-\mu)^2+\Delta^2}}\tau_z\sigma_z + \frac{\frac{k^2}{2m}-\mu}{\Delta}\tau_x\sigma_z+\frac{(\frac{k^2}{2m}-\mu)^2}{\Delta\sqrt{(\frac{k^2}{2m}-\mu)^2+\Delta^2}}i\tau_y\sigma_z
\eea

\noi Therefore,
\vspace{-12pt}
%\begin{adjustwidth}{-4.6cm}{0cm}
\bea
&&<\Delta_{p_z}> \propto \sum_{k} k  \frac{\Delta}{\sqrt{(\frac{k^2}{2m}-\mu)^2+\Delta^2}}\left (n_F(E_{+,-})-n_F(E_{+,+}) - [n_F(E_{-,-}) - n_F(E_{-,+})] \right ) \no \\
&&<\Delta_{p_z}> \propto \sum_{k} k  \frac{\Delta}{\sqrt{(\frac{k^2}{2m}-\mu)^2+\Delta^2}}\sum_{s,\pm}\mp s n_F(E_{s,\pm})
\label{Eq:4}
\eea
% \end{adjustwidth}
 \noi and apparently the $<\Delta_{p_z}>$ correlations emerge from the imbalance between singlet pairing in the $k\uparrow,-k\downarrow$ channel $<\psi_{k,\uparrow}^{\dag}\psi_{-k,\downarrow}^{\dag}+\psi_{-k,\downarrow}\psi_{k,\uparrow}> \propto n_F(E_{+,-})-n_F(E_{+,+})$ and in the $k\downarrow,-k\uparrow$ channel  $<\psi_{-k,\uparrow}^{\dag}\psi_{k,\downarrow}^{\dag}+\psi_{k,\downarrow}\psi_{-k,\uparrow}> \propto n_F(E_{-,-}) - n_F(E_{-,+})$. From \mbox{Equation~(\ref{Eq:4})} it is straightforward that, for $h=0$, $<\Delta_{p_z}> =0$ since, in this case, $n_F(E_{-,+}) = n_F(E_{+,+})$ and $n_F(E_{-,-}) = n_F(E_{+,-})$, and therefore $\sum_{s,\pm}\mp s n_F(E_{s,\pm})=0$. Moreover, for $J=0$ expression $\sum_{s,\pm}\mp s n_F(E_{s,\pm})$ is even in momenta, and therefore $<\Delta_{p_z}> =0$, due to the multiplication with momenta $k$. Thus, $p$- wave correlations are induced only when both $h \neq 0$ and $J \neq 0$.
 
 {For the other two components of the $d$ vector,  $d_y=(0,1,0)$ and  $d_x=(1,0,0)$, there are no $p$-wave correlations induced. Particularly,} for $d_y=(0,1,0)$ we get
\bea
<\Delta_{p_y}> \propto \sum_{k}\sum_{m} k n_{F}(E_m)[U(k)^{\dag}\tau_y\sigma_0 U(k)]_{mm}
\eea

\noi where 
\bea
[U(k)^{\dag}\tau_y\sigma_0 U(k)]= \frac{\Delta}{\sqrt{(\frac{k^2}{2m}-\mu)^2+\Delta^2}}\tau_z\sigma_y + \frac{\frac{k^2}{2m}-\mu}{\Delta}\tau_x\sigma_y+\frac{(\frac{k^2}{2m}-\mu)^2}{\Delta\sqrt{(\frac{k^2}{2m}-\mu)^2+\Delta^2}}i\tau_y\sigma_y
\eea

\noi Therefore,
\vspace{6pt}
\bea
<\Delta_{p_y}> =0
\eea
 
 \noi since  matrix $[U(k)^{\dag}\tau_y\sigma_0 U(k)]$ has no diagonal term. Similarly, for $d_x=(1,0,0)$ we obtain
\bea
<\Delta_{p_x}> \propto \sum_{k}\sum_{m} k n_{F}(E_m)[U(k)^{\dag}\tau_x\sigma_z U(k)]_{mm}
\eea

\noi where 
\bea
[U(k)^{\dag}\tau_x\sigma_x U(k)]= \frac{\Delta}{\sqrt{(\frac{k^2}{2m}-\mu)^2+\Delta^2}}\tau_z\sigma_x + \frac{\frac{k^2}{2m}-\mu}{\Delta}\tau_x\sigma_x+\frac{(\frac{k^2}{2m}-\mu)^2}{\Delta\sqrt{(\frac{k^2}{2m}-\mu)^2+\Delta^2}}i\tau_y\sigma_x
\eea

\noi Therefore,
\bea
<\Delta_{p_x}> =0
\eea
 
 \noi since, again, matrix $[U(k)^{\dag}\tau_x\sigma_z U(k)]$ has no diagonal terms. Thus, we conclude that only $p$-wave correlations with $<\bm{d} \parallel \bm{h}>$ emerge, due to the imbalance among the two spin configurations of interband conventional pairing.
 
%%%%%%%%%%%%%%%%%%%%%%%%%%%%%%%%%%%%%%%%%%
\section{Engineering Majorana Zero Modes without Spin-Orbit Coupling}

\subsection{Intraband \emph{p}-Wave Superconductivity in the Ferromagnetic Wire}

As demonstrated in the previous section, triplet $p$-wave correlations are induced in a conventional superconductor when a supercurrent and an exchange or Zeeman field are applied.  However, topological superconductivity and Majorana zero modes cannot be engineered in a superconducting wire simply by applying a supercurrent and a magnetic field. The triplet $p$-wave correlations induced in this setup still connect different energy bands of the wire, while MZMs require intraband or effectively spinless superconductivity. Nevertheless, when these correlations mediate to a material with magnetisation perpendicular to the exchange field induced in the superconductors, an intraband superconducting state that hosts MZMs is realised.  

Considering the particular setup presented in Figure \ref{fig1}, the induced $\Delta_p$ correlations in the SCs correspond to Cooper pairs of electrons with opposite spin, i.e., $p_z \equiv \tau_x\sigma_x$ correlations. These correlations, for which $\bm{d}=d_z$, mediate in the ferromagnet with polarisation along the $x$-axis, i.e., perpendicular to the $\bm{d}$ vector of the induced correlations, described by the following Hamiltonian   
\bea
 H_{FM}=\int dx \Psi^{\dag}(x) \biggl( [\frac{\partial_x^2}{2m_{FM}}-\mu_{FM}]\tau_z  + h_{{FM}}\tau_z\sigma_x {+ \Delta_{p}\partial_x\tau_x\sigma_x} \biggl ) \Psi(x)\,. \label{Eq:5}
\eea

\noi The energy bands of the FM wire derive from the following transformation 
\bea
\Psi'=U\Psi, \qquad  U= \left ( \begin{array}{cccc} 0 & 0 & 1 & 1 \\ 0 & 0 & -1 & 1 \\ -1 & 1 & 0 & 0 \\ 1 & 1 & 0 & 0 \end{array} \right  )
\eea

\noi where $U$ is the matrix that diagonalises the Hamiltonian $H_{FM}$. In this eigenbasis, the induced $p_z$ triplet correlations are expressed as
\bea
\Delta_{p} \propto \tau_x\sigma_0 \label{Eq:6}
\eea

\noi From Equation (\ref{Eq:6}), it  becomes apparent that, in the  eigenbasis of the FM, the induced by proximity $p$-wave field $\Delta_{p}$ pairs electrons from the same energy band, i.e., we obtain intraband pairing. The Hamiltonian Equation (\ref{Eq:5}) upon the $U$ transformation acquires the~form 
\vspace{6pt}
\bea
 H'_{FM}=\int dx \Psi'^{\dag}(x) \biggl( [\frac{\partial_x^2}{2m_{FM}}-\mu_{FM}]\tau_z  + h_{FM}\tau_z\sigma_z + \Delta_p\partial_x\tau_x\sigma_0  \biggl ) \Psi'(x)\,. \label{Eq:7}
\eea

\noi which can be reduced into two copies of the Kitaev's Hamiltonian
\bea
 H'_{FM,s}=\int dx \Psi_{s}'^{\dag}(x) \biggl( [\frac{\partial_x^2}{2m_{FM}}-\mu_{FM,s}]\tau_z + \Delta_p\partial_x\tau_x  \biggl ) \Psi_{s}'(x)\,. \label{Eq:8}
\eea

\noi where $\Psi_{s}'^{\dag}=(\psi_{s}^{\dag},\psi_{s}\up)$, $s=\pm$ the band index and $\mu_{FM,s}=\mu_{FM}+sh_{FM}$.

Based on the above analysis, it becomes clear that the optimal configuration for the emergence of MZMs in the particular platform, is that of anti-parallel supercurrents in the SCs and anti-parallel magnetisations in the FIs. This configuration guarantees that no supercurrent or exchange fields parallel to the magnetisation of the FIs are induced in the FM wire. Thus, the $d$-vector of the induced $p$-wave superconducting correlations remains perpendicular to the magnetisation of the FM wire and an intraband superconducting state is realised. However, as we will elaborate in a future work, MZMs are robust against significant deviations from this optimal configuration for the supercurrents and the FIs magnetisations. 

\subsection{Topological Criteria for the Ferromagnetic Wire}

Based on Hamiltonian Equation (\ref{Eq:7}) describing the FM wire, we derive, in this section, the criteria regarding the exchange field $h_{FM}$ and the chemical potential $\mu_{FM}$ for which MZMs emerge localised at the edges of the FM.  To this aim, we consider periodic boundary conditions and transform Hamiltonian Equation (\ref{Eq:7}) in momenta $k$ space.  Introducing the spin-dependent Nambu spinor {$\Psi_{k}=(\psi_{k,s}\up,\psi_{k,s'}\up,\psi_{-k,s'}^{\dag},\psi_{-k,s}^{\dag})^{T}$, where $\psi_{k,s}^{(\dag)}$ are the operators destroying (creating) an electron with momenta k and spin s} and the Pauli matrices $\bm{\tau}$ for the particle-hole  and $\bm{\sigma}$ for the spin space, the Hamiltonian of the FM acquires the form 
\bea
{\cal H}=\sum_k \Psi_k^{\dag} \left [(2t\cos{k}+\mu_{FM})\tau_z + h_{FM}\tau_z\sigma_z+ \Delta\tau_y\sigma_y + \Delta_p\sin{k}\tau_x\sigma_0   \right ] \Psi_k^{\dag} \,, \label{Eq:9}
\eea 
where we  also considered an induced by proximity conventional superconducting field $\Delta$. We remark that Hamiltonian Equation (\ref{Eq:9}) is expressed in lattice instead of continuum space, in order to connect with the numerical results presented in the next section. In the particular case, the reality conditions,  $\Theta H(k)\Theta^{-1}=H(-k)$ and  $\Xi H(k)\Xi^{-1}=-H(-k)$ where $\Theta$ and $\Xi$ anti-unitary operators \cite{ref-journal39}, are satisfied for  $\Xi=\tau_x{\cal K}$ with $\Xi^2=I$ and  $\Theta =i\tau_z\sigma_z{\cal K}$ with $\Theta^2=I$. In addition,  a unitary chiral symmetry operator ${\cal S}=\Theta \cdot \Xi=\tau_y\sigma_z$ anticommutes with the Hamiltonian. 

Thus, the system belongs to the BDI symmetry class, characterised by an integer $\Bbb{Z}$ topological invariant, and Hamiltonian Equation (\ref{Eq:9}) acquires a block off-diagonal form ${\cal H}'(k)=\left (\begin{array}{cc} 0 & {\cal A}(k) \\ {\cal A}(k)^{\dag} & 0 \end{array} \right)$ in the eigenbasis of the chiral operator  ${\cal S}$. Therefore, by applying the transformation $U_s=i[\tau_x+i\tau_y-\tau_z]\sigma_y+[\tau_x-i\tau_y+\tau_z]\sigma_x$, we obtain ${\cal A}(k)=h_{FM}\sigma_z-[2t\cos{k}+\mu_{FM}+i\Delta_p\sin{k}]+i\Delta\sigma_x$. The topological invariant of the Hamiltonian is defined as the winding number ${\cal N}=\frac{-i}{2\pi}\int_{k=0}^{k=2\pi}\frac{dz(k)}{z(k)}$ \cite{ref-journal40} , where $z(k)$ the unimodular complex number defined as $z(k)=Det(A(k))/|Det(A(k))|$. For the particular system, we~find 

 \bea
&&\Re z(k)=\left [(2t\cos{k}+\mu_{FM})^2-(h_{FM})^2 + \Delta^2 + \Delta_p^2\sin^2{k} \right ] \no \\
&&\Im z(k)=2\left [(2tcosk+\mu_{FM})p_x \right]\sin{k} \,.
 \eea
 
% \noi Topological phase transitions occur when the Hamiltonian acquires zero eigenvalues. The quasiparticles eigenergy spectrum of the particular Hamiltonian derives from the following equation
 
% \bea
%&&E(k)=\pm \sqrt{h_z^2+ \Delta^2+[2t\cos(k)+\mu]^2+p_x^2\sin^2(k) + \pm %2\sqrt{B}} \no \\
%&&B=\Delta^2(p_x^2\sin^2(k)+h_z^2)+[(2t\cos(k)+\mu)h_z]^2 \label{eq::HE}
% \eea

\noi  Since within each topological phase, the $\theta(k)$ is a continuous function of momentum k, the maximum possible value of the winding number relates to the number n of roots $\Im Det(A(k))$ within the interval $[0,2\pi)$, $max|{\cal N}|=n/2$. Notice that because $\Im Det(A(k)) =- \Im Det((A(-k))$, number n is always even. Next, we distinguish the two following cases. 

The first case holds for $\left |\mu \right|>2t$. In this case,  $\Im Det(A(k))=0$ only for $k=0, \pi$, and therefore the maximum absolute value of the winding number is $max|{\cal N}|=1$.  The topological ${\cal N}=1$ phase is realised when $z(k=0)z(k=\pi)=\left [\Delta^2+(2t + \mu_{FM})^2 -h_{FM}^2 \right ]$\\
$\left [\Delta^2+(2t - \mu_{FM})^2 -h_{FM}^2\right ]<0$, which results in the following condition for the chemical~potential 
\bea
|2t - \sqrt{h_{FM}^2-\Delta^2}|< |\mu_{FM}| <|2t + \sqrt{h_{FM}^2-\Delta^2}|
\eea

\noi   or equivalently for the exchange field 
\bea
\sqrt{\Delta^2+(2t-\mu_{FM})^2}<|h_{FM}|<\sqrt{\Delta^2+(2t+\mu_{FM})^2}\,.
\eea 

The second case holds for $\left |\mu_{FM} \right|<2t$. When this condition holds, $\Im Det(A(k))=0$ is realised also for another pair of momenta $k^*=\pm cos^{-1}\left (\left |\frac{\mu_{FM}}{2t} \right | \right )$. However, since for $\sqrt{\Delta^2+(2t-\mu_{FM})^2}<|h_{FM}|<\sqrt{\Delta^2+(2t+\mu_{FM})^2}$, $z(k=0)z(k=\pi)<0$, it is straightforward that $|{\cal N}|=1$ within these limits, even in this case. In general, considering $\Delta_p<<\Delta$,  a $|{\cal N}|=2$ topological phase can be realised for 
\bea
\Delta \lesssim |h_{FM}|<\sqrt{\Delta^2+(2t-\mu_{FM})^2}
\eea

\noi for which $z(k=0)z(k=\pi)>0$ and $z(k=k^*)z(k=\pi)<0$.

\subsection{Numerical Calculations for the FI-SC-FM-SC-FI Heterostructure}

In order to verify the above arguments,  we employ the following lattice Bogoliubov de Gennes equation,
\bea
{\cal H} = [H_{l,SC}+H_{l,tc}+H_{FM}+H_{r,tc}+H_{r,SC}] \label{Eq:10}
\eea

\noi  which describes the FI-SC-FM-SC-FI heterostructure presented in Figure \ref{fig1}, with
\bea
H_{a,SC}= \sum_{\bm{i},\bm{j}}\Psi_{a,\bm{i}}^{\dag}\left [ (tf_{\bm{i},\bm{j}}+\mu_{SC})\tau_z + h_{a,SC}\tau_z\sigma_z + \Delta\tau_y\sigma_y + J_{a} \cdot \bm{g}_{\bm{i},\bm{j}}\right ] \Psi_{a,\bm{j}}^{\dag} \label{Eq:11}
\eea

\noi the Hamiltonian of the superconductor, where $a=l,r$ stand for left and right SC with $h_{l,SC}=-h_{r,SC}$, $J_{l}=-J_{r}$  and
\bea
H_{FM}= \sum_{\bm{i},\bm{j}}\Phi_{\bm{i}}^{\dag} \left [(tf_{\bm{i},\bm{j}}+\mu_{FM})\tau_z + h_{FM} \tau_z\sigma_x \right ]\Phi_{\bm{j}}^{\dag} \label{Eq:12}
\eea

\noi the Hamiltonian of the FM, where $\Psi_{a,\bm{i},SC}^{\dag} = (\psi_{a,\uparrow,\bm{i}}^{\dag},\psi_{a,\downarrow,\bm{i}}^{\dag},\psi_{a,\uparrow,\bm{i}}\up,\psi_{a,\downarrow,\bm{i}}\up)$ and 
$\Phi_{\bm{i},FM}^{\dag}$\linebreak $= (\phi_{\uparrow,\bm{i}}^{\dag},\phi_{\downarrow,\bm{i}}^{\dag},\phi_{\uparrow,\bm{i}}\up,\phi_{\downarrow,\bm{i}}\up)$ {with $\psi_{a,s,\bm{i}}^{(\dag)}$ and $\phi_{s,\bm{i}}^{(\dag)}$ the operators destroying (creating) an electron with spin s of the $a=l,r$ SC and FM, respectively, at lattice site  $\bm{i}$,} $\bm{\tau}$ and $\sigma$ the Pauli matrices acting on particle-hole and spin space respectively, $f_{\bm{i},\bm{j}}=\delta_{\bm{j},\bm{i} \pm \hat{x}} + \delta_{\bm{j},\bm{i} \pm \hat{y}}$ and $\bm{g}_{\bm{i},\bm{j}}=(\pm i\delta_{\bm{j},\bm{i}\pm \hat{x}}, \pm i \delta_{\bm{j},\bm{i}\pm \hat{y}})$ the even and odd in spatial inversion, respectively, functions connecting nearest neighbours lattice points. Moreover, $t$ is the hopping integral and $J$ the supercurrent term. The proximity of the SCs to the FIs is modelled by the introduction of an exchange field term in the $H_{SC}$ Equation (\ref{Eq:11}). The SCs are connected with the FM wire through the following Hamiltonian 
\bea
H_{a,tc}= \sum_{\bm{i},\bm{j},s}t_c \psi_{a,{\bm{i}},s}^{\dag} \phi_{{\bm{j}},s}\up + c.c. \,. \label{Eq:13}
\eea

\begin{figure}[h]
\begin{tabular}{cc}
\includegraphics[scale=0.2]{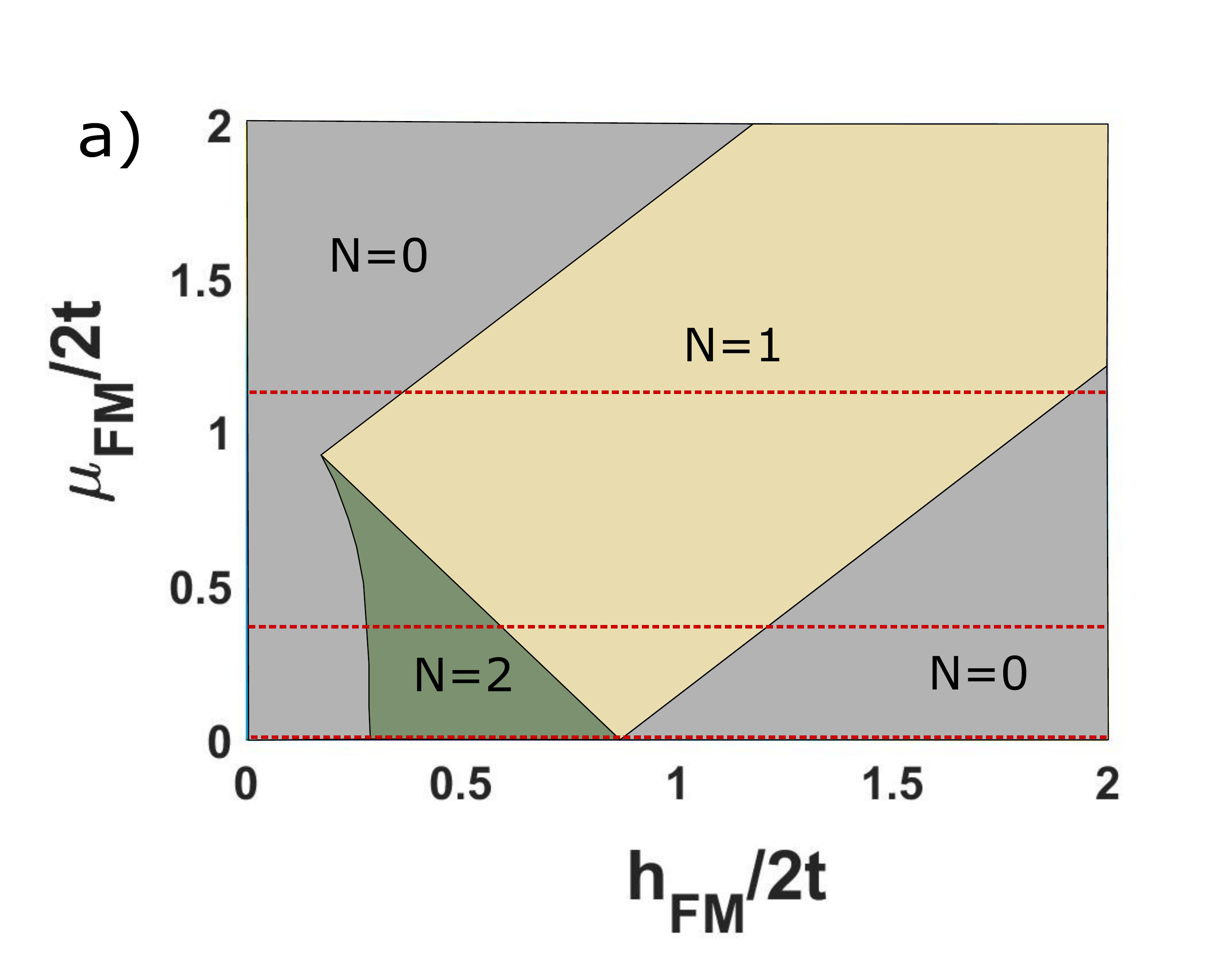}
\includegraphics[scale=0.25]{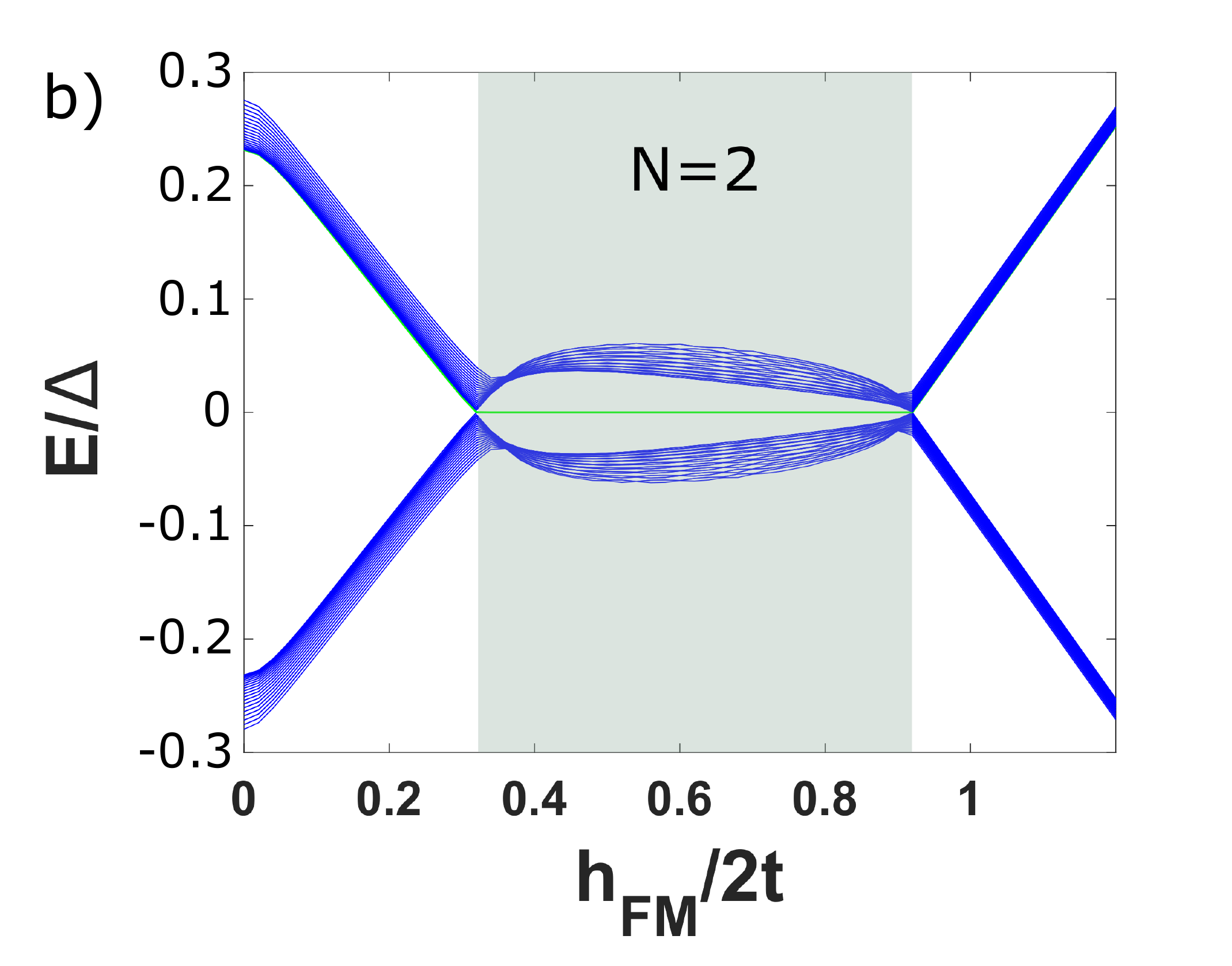}\\
\includegraphics[scale=0.25]{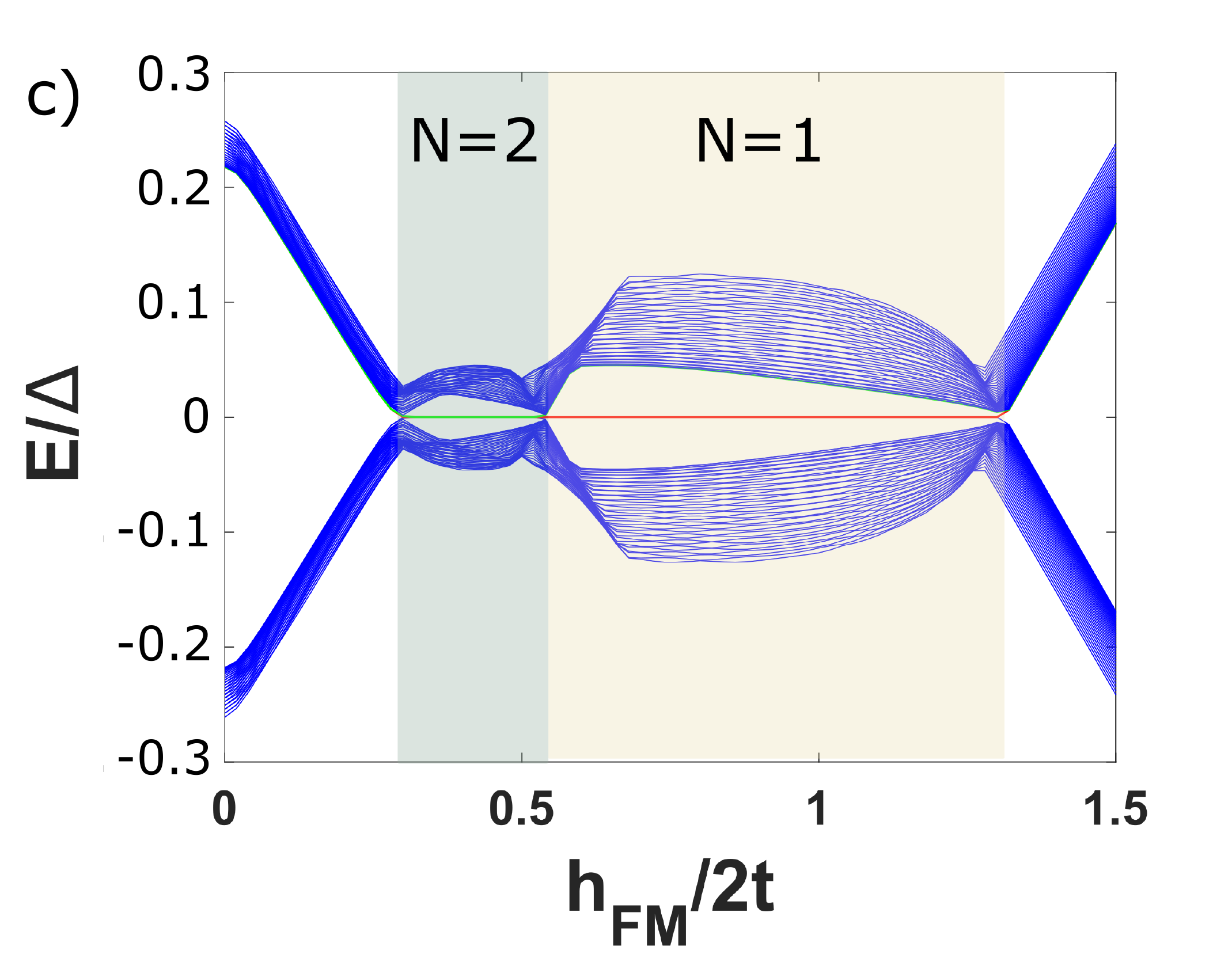}
\includegraphics[scale=0.25]{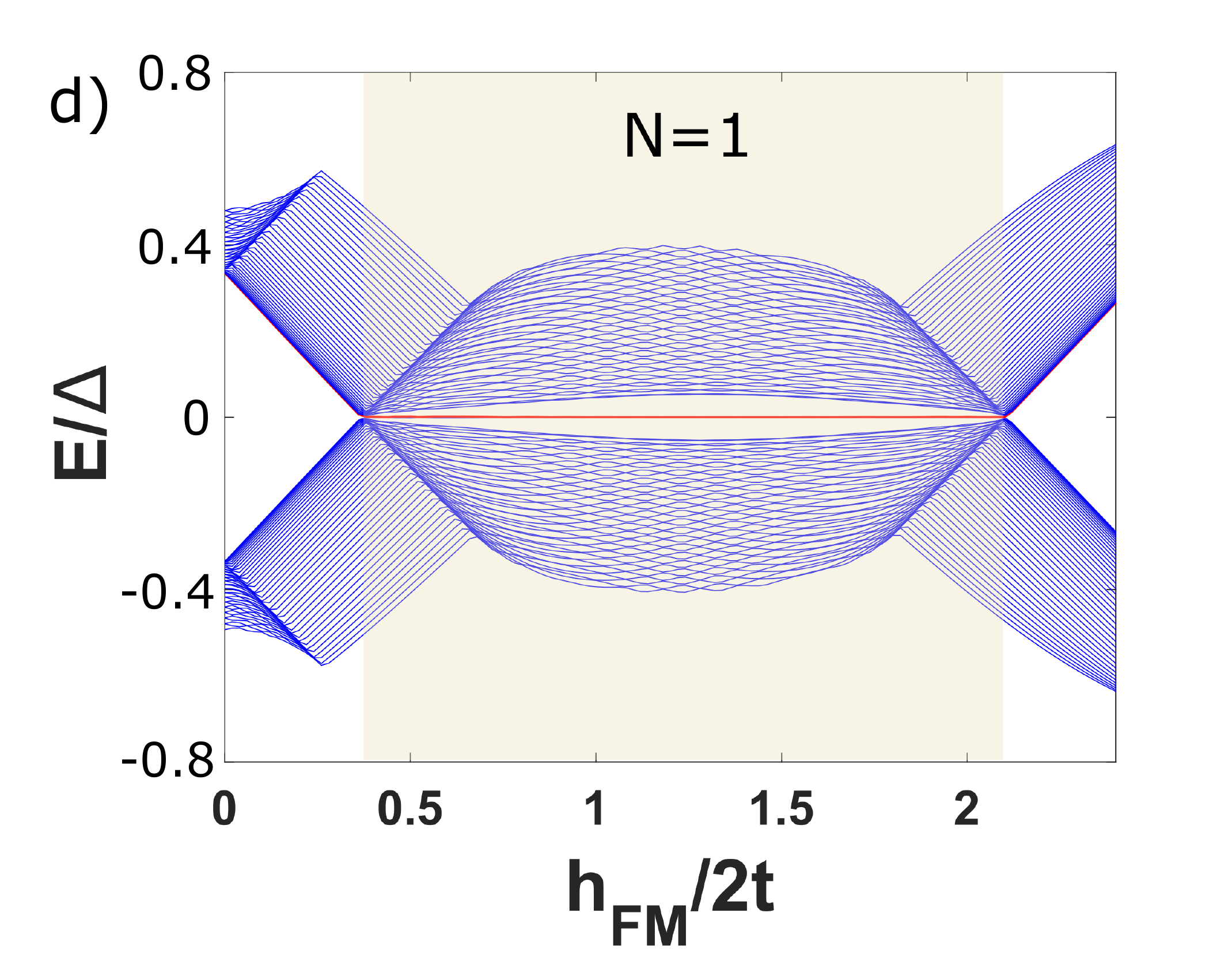}
\end{tabular}
\caption{ For $\Delta/2t=1$, $\mu_{SC}/2t=0$,  $J/\Delta=0.2$ and $h_{SC}/\Delta=0.2$, (\textbf{a}) topological phase diagram with respect to the chemical potential $\mu_{FM}$ and the exchange field $h_{FM}$ of the ferromagnet.  The low energy spectra with respect to $h_{FM}$ for (\textbf{b}) $\mu_{FM}/2t=0$, (\textbf{c}) $\mu_{FM}/2t=0.4$ and (\textbf{d}) for $\mu_{FM}/2t=1.2$ denoted by red dashed lines in figure (\textbf{a}). The numerical results verify the topological criteria presented in the previous section.  \label{fig2}}
\end{figure}

\noi  In general, we consider $t_c=t$ for simplicity. MZMs are anticipated to emerge localised at the edges of the wire for $|2t - \sqrt{(h_{FM})^2-\Delta'^2}|< |\mu_{FM}| <|2t + \sqrt{(h_{FM})^2-\Delta'^2}|$ where $\Delta'$ is the induced by proximity { conventional} superconducting field over the FM. In Figure \ref{fig2}a we present the phase diagram for the setup of Figure \ref{fig1} considering $\Delta/2t=1$, $\mu_{SC}/2t=0$, $\bm{J}/2t=0.2$ and $h_{SC}/2t=0.2$. Based on this diagram, we observe that for $\mu_{FM}<2t$, the induced singlet superconducting field over the FM is $\Delta'/2t \simeq 0.25$ while the hopping term along the wire is normalised to $t'\simeq 0.86t$, due to the coupling of the wire to the SCs \cite{ref-journal41}. 

For  $\mu_{FM}>2t$, again, the hopping term along the wire is normalised to $t'\simeq 0.86t$ and the black lines $h_{FM}/2t \simeq \mu_{FM}/2t - 2t'/2t$ and $h_{FM}/2t \simeq \mu_{FM}/2t + 2t'/2t$ define the region of parameters for which ${\cal N} =1$ and a pair of MZMs emerge at the edges of the FM wire. Notice that the particular phase diagram is in accordance with the topological criteria for the ${\cal N}=1$ and ${\cal N}=2$ topological phases presented in the previous section. { However, we remark that the particular results correspond to a fixed conventional order parameter, which has not been self-consistently determined over the SC regions. 
	
	Therefore, the effect of finite momentum Cooper pairs, which, in principle, can emerge due to the presence of the supercurrent and the exchange field in the conventional superconductor, on the phase diagram (Figure \ref{fig2}), has not been investigated and will be examined in a future work.} 

\vspace{-12pt}

\subsection{Local Density of States}

The pursuit of functional platforms for topological quantum computation, requires the experimental verification of the theoretical proposals for engineering topological superconductivity, by detecting the emergent MZMs. A characteristic signature of MZMs in conductance spectroscopy measurements is the quantized zero-bias tunnelling conductance. Emanating from an Andreev reflection resonant at zero energy, the zero bias conductance in the presence of a single MZMs pair equals $\frac{2e^2}{h}$, i.e., double the conductance quantum, irrespective of the tunnelling barrier \cite{ref-journal42,ref-journal43}. 

Several tunnelling experiments have demonstrated such zero bias conductance peaks and therefore provided supporting evidence for the emergence of MZMs in the semiconductor nanowires \cite{ref-journal44}, the topological insulator/conventional superconductor heterostructures~\cite{ref-journal45,ref-journal46}, the magnetic adatoms embedded in conventional superconductor \cite{ref-journal47}, the heavy metal surfaces \cite{ref-journal48}, the quantum anomalous Hall insulator \cite{ref-journal49}, the iron-based superconductors \cite{ref-journal50} and the planar Josephson junctions \cite{ref-journal51} platforms.  However, the results are far from being conclusive, since in each case several deviations from the corresponding theoretical predictions are observed.

Since differential conductance measurements in scanning tunnelling spectroscopy experiments are proportional to the local density of states of the sample, when the scanning tunnelling junction can be approximated by a point contact and the tunnelling amplitude is small, we calculate the local density of states at characteristic positions in the FM wire in order to link our results with relevant experiments. To this end, we solve the following eigenvalue equation
\bea
\hat{H}_{\bm{i},\bm{j}}
\left(
\begin{array}{c}
u_{n,\uparrow}(\bm{j}) \\
u_{n,\downarrow}(\bm{j}) \\
v_{n,\uparrow}(\bm{j}) \\
v_{n,\downarrow}(\bm{j}) 
\end{array}
\right)=E_{n}
\left (
\begin{array}{c}
u_{n,\uparrow}(\bm{i}) \\
u_{n,\downarrow}(\bm{i}) \\
v_{n,\uparrow}(\bm{i}) \\
v_{n,\downarrow}(\bm{i}) 
\end{array}
\right)
\eea 

\noi where $\hat{H}_{\bm{i},\bm{j}}$ the Hamiltonian matrix of the FI-SC-FM-SC-FI heterostructure (Equation (\ref{Eq:10})) and $\left(
\begin{array}{cccc}
u_{n,\uparrow}(\bm{j}) &
u_{n,\downarrow}(\bm{j}) &
v_{n,\uparrow}(\bm{j}) &
v_{n,\downarrow}(\bm{j}) 
\end{array}
\right)^T$ the eigenstate corresponding to eigenenergy $E_n$. Based on the eigenstates and eigenvalues of the Hamiltonian we calculate the local density of states for spin up and down components according to the following equations
\bea
&&N(\bm{i},\omega)_{\uparrow}=-\frac{1}{\pi}Im\hat{G}^{R}_{\bm{i},\bm{i}, \omega,\uparrow\uparrow}= \sum_{n} \left[ |u_{n,\uparrow}(\bm{i})|^2\delta(\omega-E_n)\right ]
\eea
\bea
&&N(\bm{i},\omega)_{\downarrow}=-\frac{1}{\pi}Im\hat{G}^{R}_{\bm{i},\bm{i}, \omega,\downarrow\downarrow}= \sum_{n} \left[ |u_{n,\downarrow}(\bm{i})|^2\delta(\omega-E_n)\right ]
\eea

\noi where $\hat{G}^{R}_{\bm{i},\bm{j},\omega}=\frac{1}{\omega+i\eta-\hat{H}_{\bm{i},\bm{j}}}$ the retarded Green's function. In order to account for the presence of impurities in the system, we consider a scattering time $\tau$,  which corresponds to a mean free path  $l=u_F\tau$, where $u_F$ the Fermi velocity. Including the scattering from impurities, the density of states  derives from equation  
\bea
&&N(\bm{i},\omega)= \frac{1}{\pi}\sum_{n} \left[ (|u_{n,\uparrow}(\bm{i})|^2+|u_{n,\downarrow}(\bm{i})|^2)\frac{\gamma}{(\omega-E_n)^2+\gamma^2}\right ]
\eea 

\noi where the Dirac function $\delta(\omega-E_n)$ has been substituted by a Cauchy--Lorentz function $\frac{1}{\pi}\frac{\gamma}{(\omega-E_n)^2+\gamma^2}$, with $\gamma=1/\tau$. Taking the clean superconductor limit, we assume that $l/\xi \simeq 1000$, where $\xi= \frac{u_F}{\pi \Delta}$ the superconducting coherence length, and therefore $\frac{\gamma}{\Delta}=\frac{\pi \xi}{l} \simeq 0.003$.

\begin{figure}[h]
\begin{tabular}{ccc}
\includegraphics[scale=0.21]{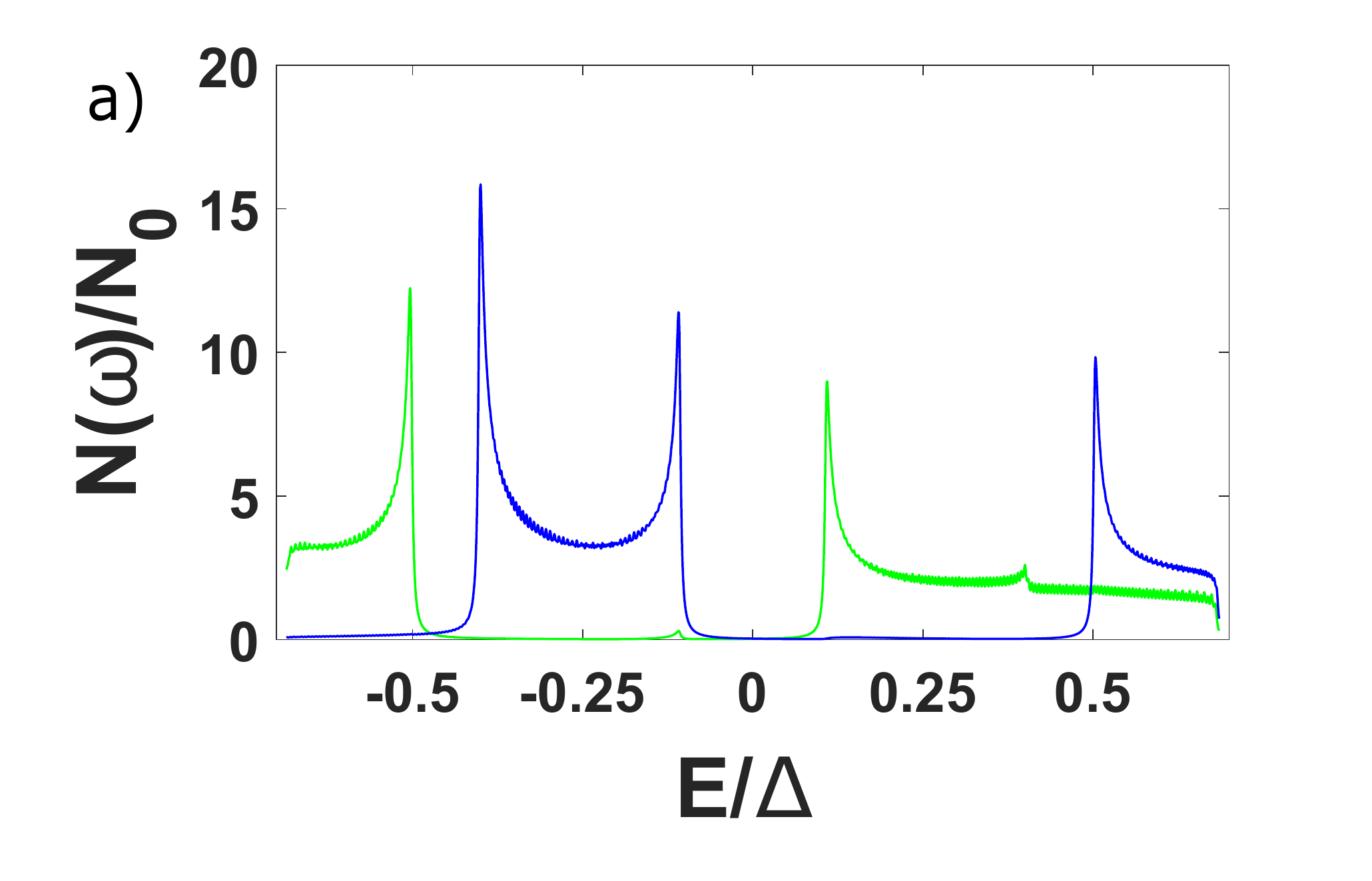}
\includegraphics[scale=0.21]{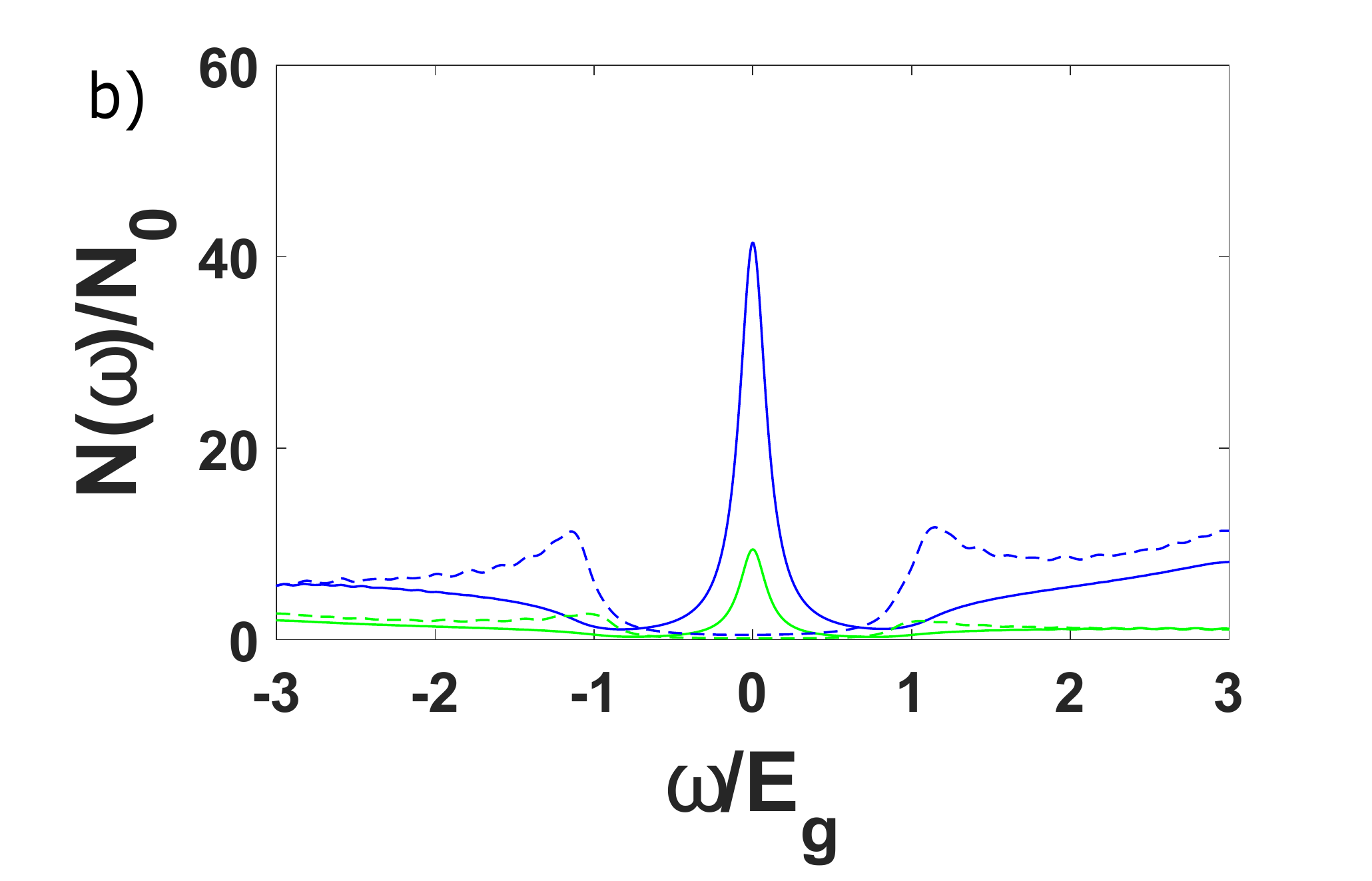}
\includegraphics[scale=0.21]{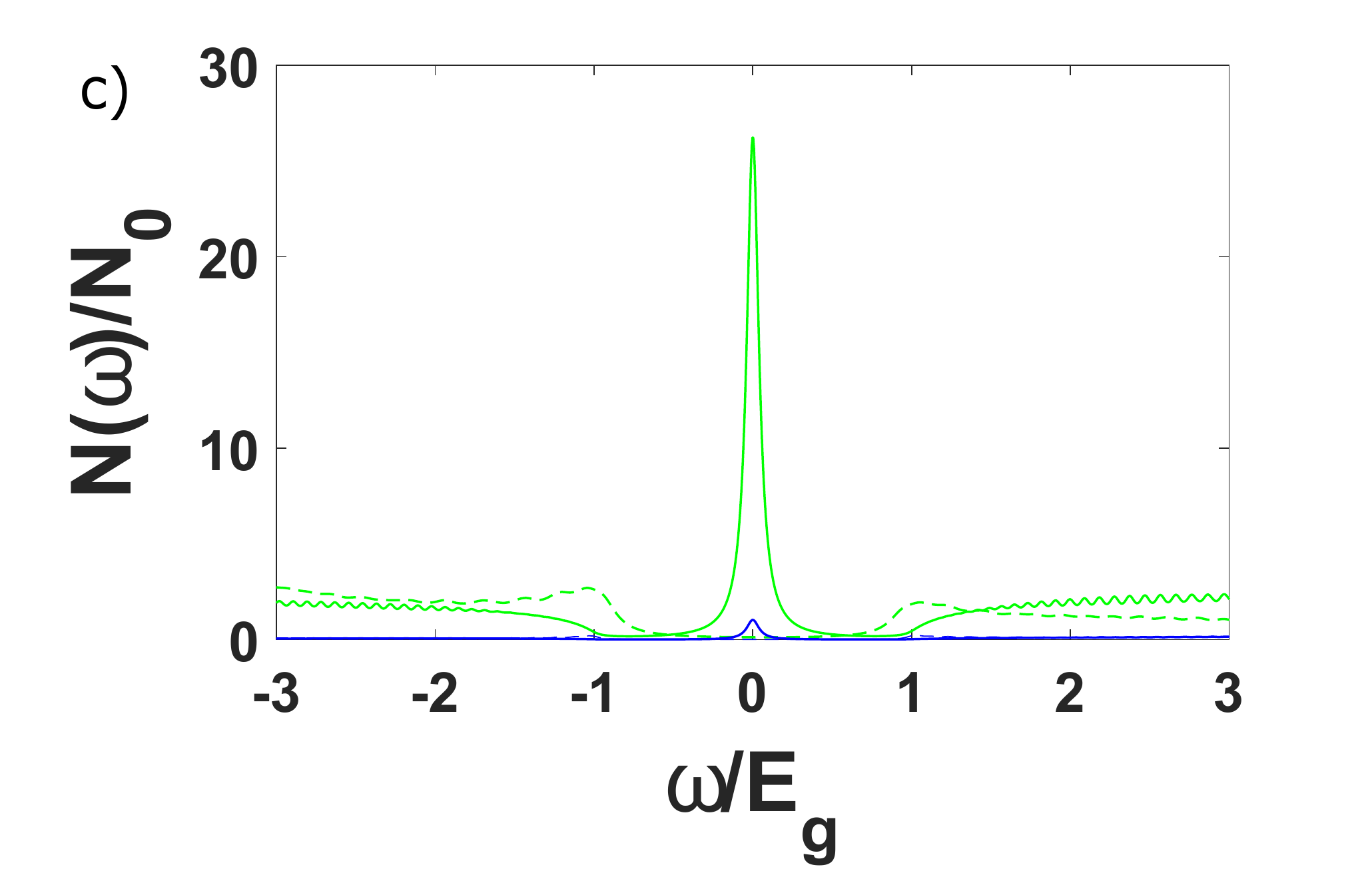}
\end{tabular}
\caption{ The local density of states %MDPI: Please revise the hyphens (-) to minus (−).
 $N(\omega)$ at the edge and the middle of the FM wire for (\textbf{a}) $h_{FM}/2t=0.18$ for which the wire is in a non-topological phase where the induced singlet pairing $\Delta'$ prevails, (\textbf{b}) for $h_{FM}/2t=0.36$ for which a topological ${\cal N}=2$ phase is realised and (\textbf{c}) for $h_{FM}/2t=0.72$ for which the wire is in a topological ${\cal N}=1$ phase. Solid and dashed lines correspond to the edge, mean value of the density at the first five lattice sites of the wire, and the middle of the wire, respectively. The blue line corresponds to the spin up, while the green line corresponds to the spin down component of the local density of states. The particular diagrams are relevant to spin-polarised scanning tunnelling microscopy (SP$-$STM) measurements tips polarised (anti$-$)parallel to the magnetisation of the FM wire.   In (\textbf{b},\textbf{c}) diagramms energy is normalised with respect to the triplet superconducting energy gap $E_g$ induced in the wire. $N_0$ is the density of states in the normal~phase. \label{fig3}}
\end{figure}

In Figure \ref{fig3}, we present the local density of states $N(\omega)$ for three values of the exchange field $h_{FM}$ of the wire considering $\mu_{FM}/2t=0.4$, which corresponds to diagram (c) of \mbox{Figure \ref{fig2}}. From Figure \ref{fig3}a we deduce that the singlet pairing field induced in the wire, due to proximity with the conventional SC, is $\Delta'/\Delta \simeq 0.25$. Since in this case $h_{FM}<\Delta'$ the FM is in a non-topological phase with ${\cal N}=0$. When $h_{FM}>\Delta'$ the wire is in a topological ${\cal N}=2$ (Figure \ref{fig3}b) or ${\cal N}=1$ (Figure \ref{fig3}c) phase, and a peak for $\omega=0$ emerges at the local density of states spectrum only at the edges of the wire, signifying the presence of localised zero energy states, the MZMs. 

For $h_{FM}/2t=0.36$ where ${\cal N}=2$ the two MZMs localised at the beginning of the wire, acquire opposite spin polarisation. The difference in the two zero energy peaks  in \mbox{Figure \ref{fig3}}b indicates the different localization of the two MZMs, due to the difference in the Fermi velocity, $u_F = 2tsin(cos^{-1}(\frac{-(\pm h_{FM}+\mu_{FM})}{2t})$, among the two spin bands. The localization length $l_M$ of the MZMs equals $l_M =\frac{u_F}{\Delta_p}$. For $h_{FM}/2t=0.72$, a single pair of MZMs with particular spin polarisation emerges localised at the FM wire as indicated from the zero energy peak, which is significant for the spin down component. However, the zero energy density of states is also finite for the other spin component, due to the finite $h_{SC}$.  

\subsection{Robustness of Emergent Majorana Zero Modes against Disorder}

In realistic experimental setups, the presence of disorder is unavoidable. On this account, we demonstrate, in this section, the robustness of the emergent MZMs against disorder in the FM wire. In particular, we consider a random on-site potential $\delta\mu_{FM}(x)\tau_z$ that derives from a normal distribution with mean value 0 and variance $\sigma^2$ corresponding to correlator $<\delta\mu_{FM}(x)\delta\mu_{FM}(x')>=\sigma^2\delta(x-x')$, i.e., $\delta\mu_{FM}(x)={\cal D}(0,\sigma)$. 

\begin{figure}[h]
\begin{tabular}{ccc}
\includegraphics[scale=0.21]{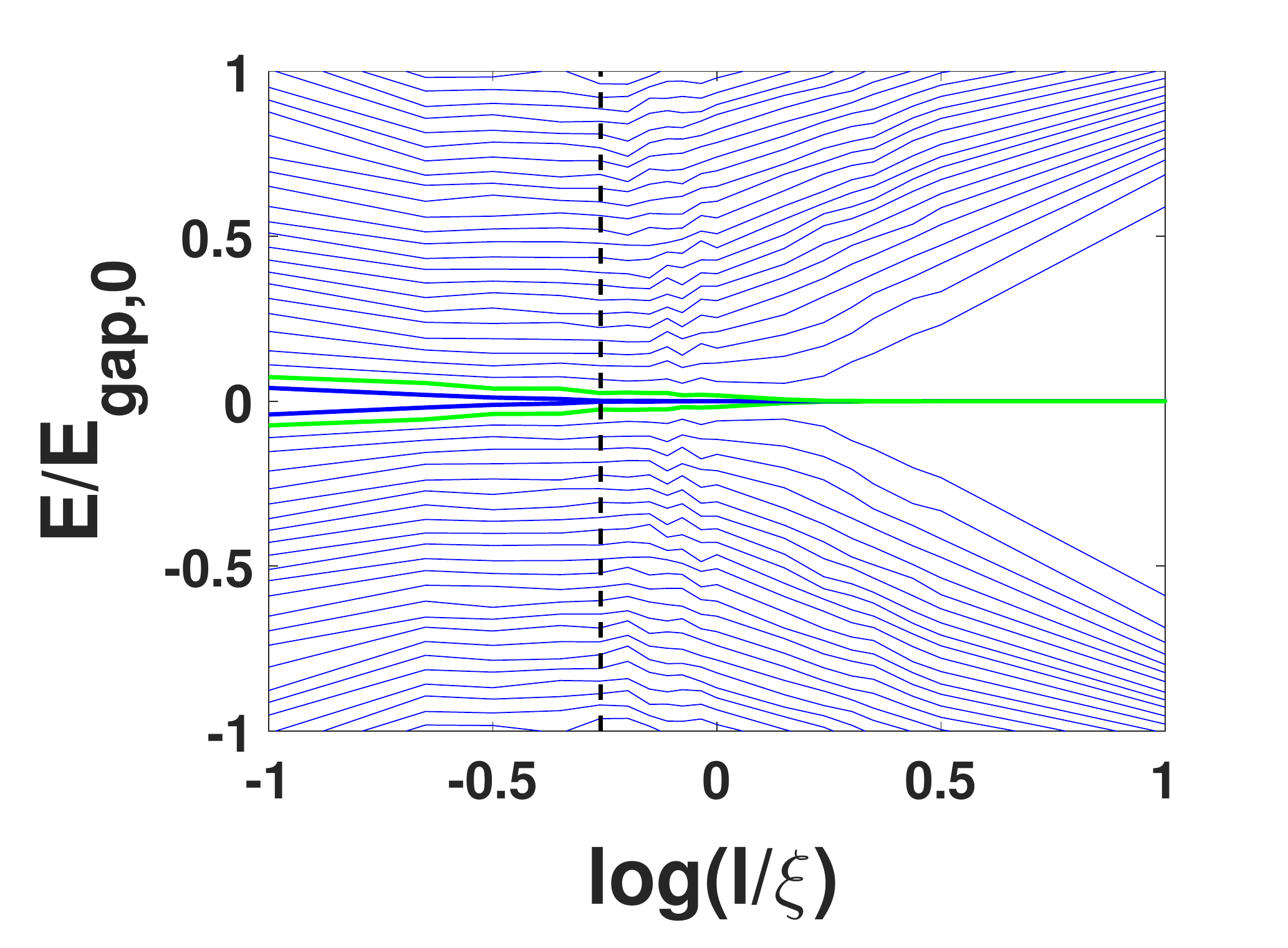}
\includegraphics[scale=0.21]{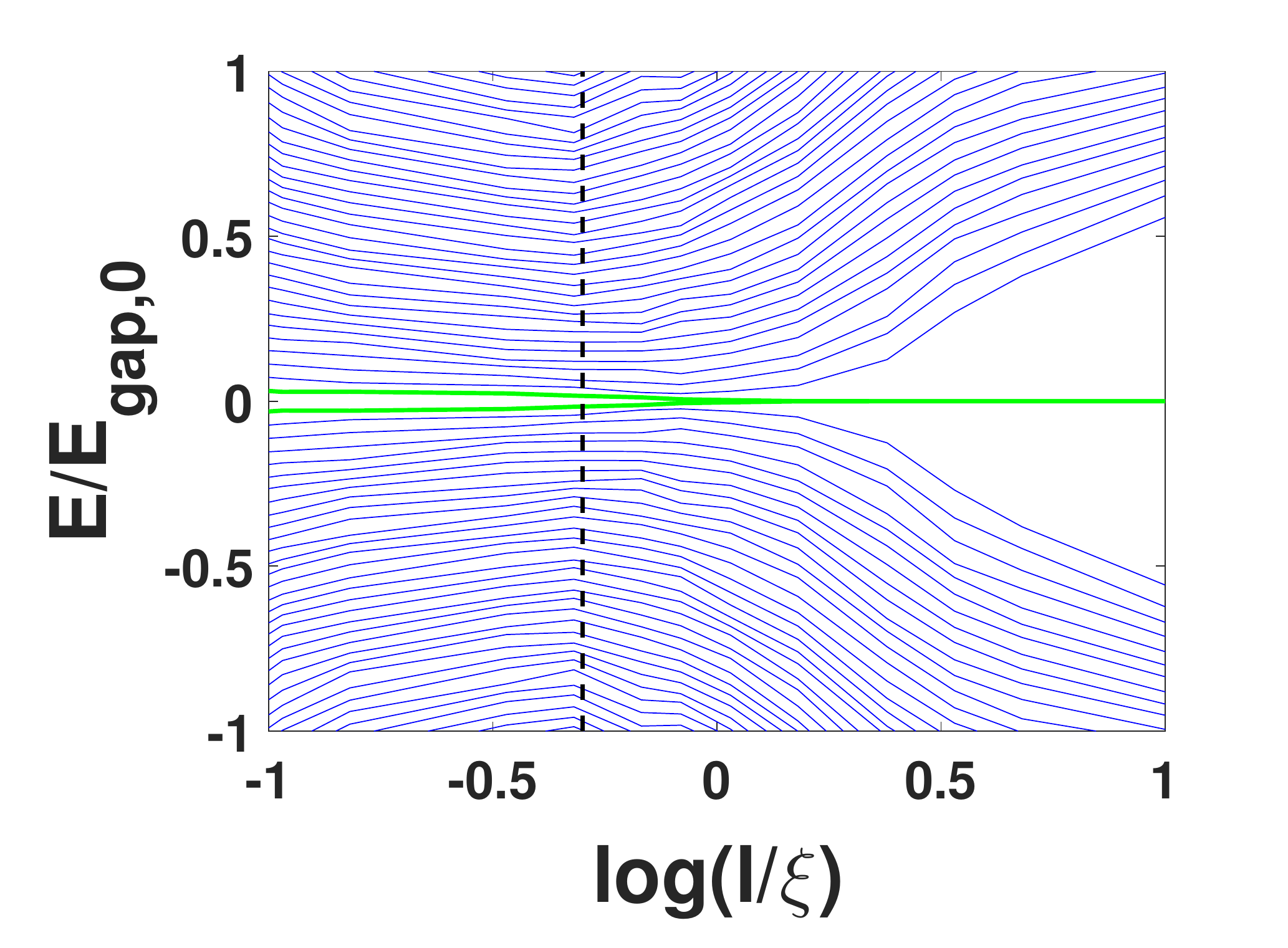}
\includegraphics[scale=0.21]{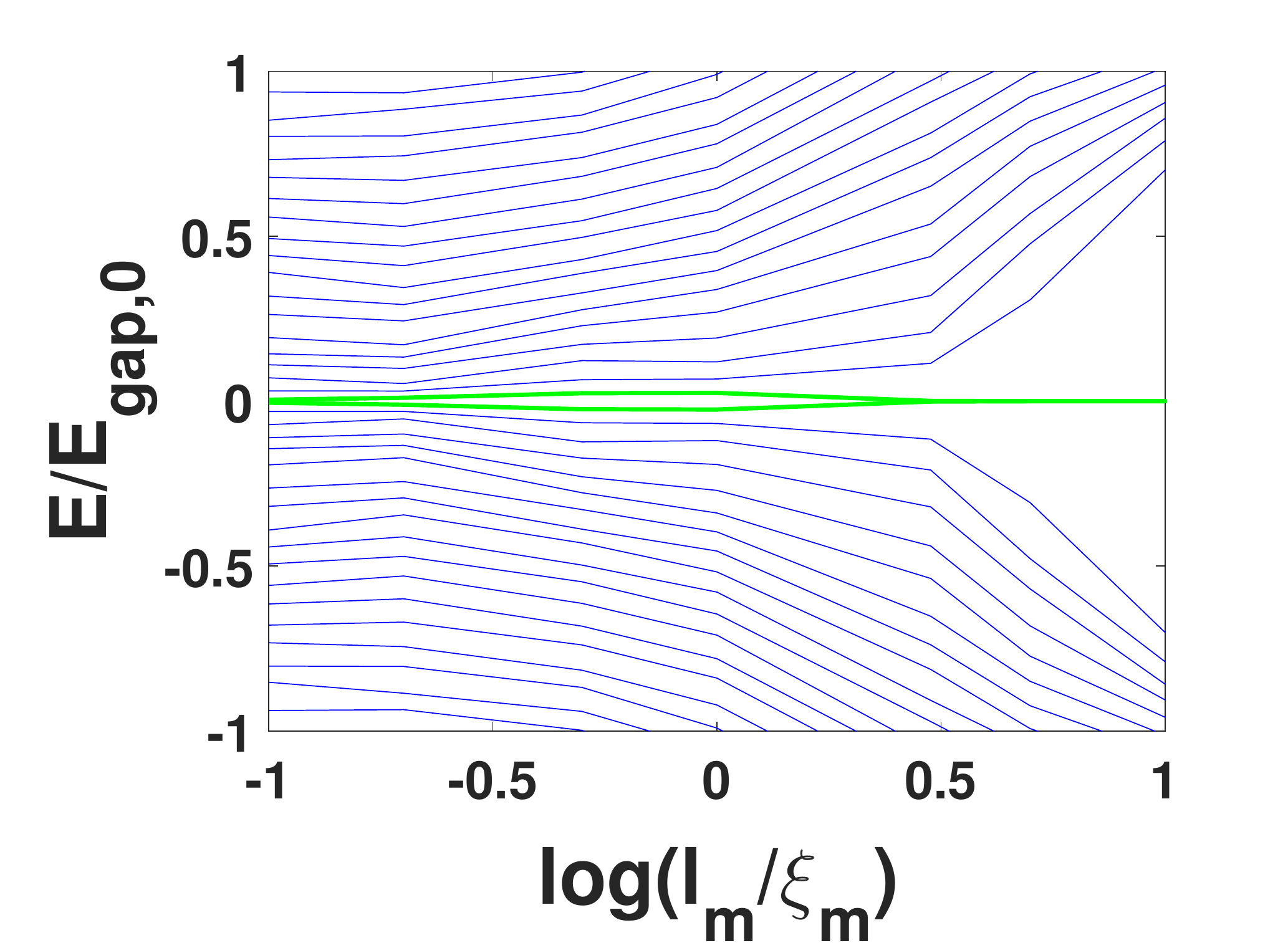}
\end{tabular}
\caption{Low energy spectrum of the FM wire with respect to the chemical %MDPI: Please revise the hyphens (-) to minus (−).
 potential disorder strength $l/\xi$ for $\mu_{FM}/2t=0.4$ and (\textbf{left}) $h_{FM}/2t=0.36$, (\textbf{middle})  $h_{FM}/2t=0.72$. The dashed black line denotes critical disorder $2l=\xi$, for which a topological phase transition occurs in $p-$wave superconducting wires. In both cases, the energy gap is reduced by $50\%$ with respect to that of a pristine wire, for $l \simeq 10\xi$. (\textbf{right}) For $h_{FM}/2t=0.72$ the low energy spectrum with respect to magnetic disorder $l_m/\xi_m$. Although the energy gap is again reduced by approximately $50\%$ for $l_m = 10\xi_m$, the MZMs appear to acquire finite energy only for $l_m<3\xi_m$. Green and blue colours for the zero energy modes correspond to spin components as in Figure \ref{fig3}. $E_{gap,0}$ is the energy gap in the pristine ferromagnetic wire.   
\label{fig4}}
\end{figure}

The mean free path $l$ along the wire is related to the variance through the following relation $l=u_F\tau=u_f^2/\sigma^2=u_F/[\pi\sigma^2N(\mu)]$ where $u_F$ the Fermi velocity and $N(\mu)=1/\pi u_F$ the density of states at the Fermi level. In the lattice model described by Hamiltonian Equation (\ref{Eq:12}), $u_F$ in the FM wire equals $u_F(\mu)=2tsin[cos^{-1}(-\mu/2t)]$ considering a common Fermi velocity for the two spin components. We define the dimensionless parameter $\kappa=\frac{l}{\xi}=\frac{\Delta_p}{\sigma^2N(\mu)}$, where $\xi=\frac{u_F}{\pi\Delta_p}$ is the coherence length of the $p$-wave superconductivity induced in the FM wire. In Figure \ref{fig4}, we present the low energy spectrum of the wire for $\mu_{FM}/2t=0.4$ and $h_{FM}/2t=0.36$ for which ${\cal N}=2$ and $h_{FM}/2t=0.72$ for which ${\cal N}=1$.

For $h_{FM}/2t=0.36$, we observe that the energy gap essentially closes for $l/\xi=2$, however, the spin up MZM remains at zero energy until $2l=\xi$, which is the critical value of disorder for $p$-wave superconducting wires \cite{ref-journal52}. We attribute the difference in the critical mean free path for the two MZMs pairs in their different localization. For $h_{FM}/2t=0.72$ the energy gap closes at $l \simeq \xi$. We remark that in both cases, the energy gap is reduced by $50\%$ with respect to that of a pristine wire when $l \simeq 10\xi$, signifying the robustness of the emergent MZMs against disorder of the chemical potential of the FM wire. 

Finally, we investigate the robustness of the emergent MZMs against disorder in the exchange field $h_{FM}$ of the FM wire. We remark that our mechanism does not depend on the structural characteristics of the FM wire, which  can therefore  also be considered as an array of ferromagnetically aligned magnetic impurities. Magnetic impurities induce in the superconductor subgap bound states, the Yu--Shiba--Rusinov states \cite{ref-journal53,ref-journal54,ref-journal55}. Depending on the hopping integral among the impurities, these subgap states form an energy band, which may lie within the superconducting gap (Shiba limit), or extent above the superconducting energy gap (wire limit) \cite{ref-journal18}. Our mechanism is valid in both limits. 

Considering this possible formation of the FM wire, it is worth examining the robustness of the emergent MZMs against disorder in the ferromagnetic alignment of the magnetic impurities and, therefore, in the corresponding orientation of the local exchange field, which otherwise is considered to have a fixed magnitude. In complete analogy to the previous case, we consider two normal distributions ${\cal D}_{\theta}=(0,\sigma_m)$ and ${\cal D}_{\phi}=(0,\sigma_m)$ for the angle $\theta$ of the local magnetisation  with respect to the $x$-axis and for the angle $\phi$ in the $y-z$~plane. 

We define a magnetic mean free path $l_m=\frac{u_F}{\sigma_m^2 N(\mu)}$ as a measure of disorder of the local exchange field and the magnetic coherence length $\xi_m=\frac{u_F}{\pi h_{FM}}$. In Figure \ref{fig4}, we present the low energy spectrum of the FM wire with respect to disorder strength $l_m/\xi_m$ only for $h_{FM}/2t=0.72$. We note that for $h_{FM}/2t=0.36$ the two pairs of MZMs interact with each other and acquire finite energy even for infinitesimal magnetic disorder, since magnetic disorder breaks the spin symmetry that prevents the two MZMs from interacting with each~other.

%%%%%%%%%%%%%%%%%%%%%%%%%%%%%%%%%%%%%%%%%%
%\section{Manipulating Majorana zero modes}

%%%%%%%%%%%%%%%%%%%%%%%%%%%%%%%%%%%%%%%%%%
\section{Conclusions}

We present a novel platform for realising Majorana zero modes based on \\
superconductor--ferromagnet heterostructures. The platform consists of a ferromagnetic wire placed among conventional superconductors, which are proximised by ferromagnetic insulators. Instead of spin-orbit coupling, the essential element of the physical mechanism underlying the proposed platform is the misalignment between the magnetisation of the ferromagnetic wire and that of the ferromagnetic insulators. 

The robustness of the emergent MZMs against impurities disorder resembles that of a $p$ superconducting wire, although the emergence of two pairs of MZMs and the observation of a zero bias conductance peak in the corresponding phase is hindered even for infinitesimal disorder in the magnetisation of the FM wire. We assert that recent developments in heterostructures fabrication renders our platform experimentally feasible, while the controllable application of the supercurrents in the superconductors provides an additional advantage with respect to other relevant proposals.

%%%%%%%%%%%%%%%%%%%%%%%%%%%%%%%%%%%%%%%%%%
\vspace{6pt} 

\abbreviations{Abbreviations}{The following abbreviations are used in this manuscript:\\

\noindent 
\begin{tabular}{@{}ll}
FM & ferromagnetic wire \\
SC & superconductor \\
FI & ferromagnetic insulator \\
MZMs & Majorana zero modes\\
TSC & topological superconductivity
\end{tabular}}

\end{paracol}
%%%%%%%%%%%%%%%%%%%%%%%%%%%%%%%%%%%%%%%%%%
% To add notes in main text, please use \endnote{} and un-comment the codes below.
%\begin{adjustwidth}{-5.0cm}{0cm}
%\printendnotes[custom]
%\end{adjustwidth}
%%%%%%%%%%%%%%%%%%%%%%%%%%%%%%%%%%%%%%%%%%
\reftitle{References}


\begin{thebibliography}{999}
% Reference 1
\bibitem[Ladd(2010)]{ref-journal1}
Ladd, T.D.; Jelezko, F.; Laflamme, R.; Nakamura, Y.; Monroe, C.; O’Brien, J.L. Quantum computers. {\em Nature} {\bf 2010}, {\em 464}, 45--53.
% Reference 2
\bibitem[Lindner(2013)]{ref-journal2}
Stern, A.; Lindner, H.N. Topological Quantum Computation—From Basic Concepts to First Experiments. {\em Science} {\bf 2013}, {\em 339}, 1179.
% Reference 3
\bibitem[Ville(2017)]{ref-journal3}
Lahtinen, V.; Pachos, J. A short introduction to topological quantum computation. {\em SciPost Phys} {\bf 2017}, {\em 3}, 021.
% Reference 4
\bibitem[Nayak(2008)]{ref-journal4}
Nayak, C.; Simon, S.; Stern, A.; Freedman, M.; Sarma, D.S. Non-Abelian anyons and topological quantum computation. {\em Rev. Mod. Phys.} {\bf 2008}, {\em 80}, 1083.
% Reference 5
\bibitem[Sato(2017)]{ref-journal5}
Sato, M.; Yoichi, A. Topological superconductors: A review. {\em Rep. Prog. Phys.} {\bf 2017}, {\em 80}, 076501.
% Reference 6
\bibitem[Kitaev(2013)]{ref-journal6}
Kitaev, A.Y. Unpaired Majorana fermions in quantum wires. {\em Physics-Uspekhi} {\bf 2001}, {\em 44}, 131.
% Reference 7
\bibitem[Ivanov(2013)]{ref-journal7}
Ivanov, D.A. Non-Abelian Statistics of Half-Quantum Vortices in \emph{p}-wave Superconductors. {\em Phys. Rev. Lett.} {\bf 2001}, {\em 86}, 268.
% Reference 8
\bibitem[Sigrist(1995)]{ref-journal8}
Rice, T.M.; Sigrist, M. Sr$_2$RuO$_4$: An electronic analogue of 3He? {\em J. Phys. Condens. Matter} {\bf 1995}, {\em 7}, L643.
% Reference 9
\bibitem[Maeno(2013)]{ref-journal9}
Maeno, Y.; Kittaka, S.; Nomura, T.; Yonezawa, S.; Ishida, K. Evaluation of spin-triplet superconductivity in Sr$_2$RuO$_4$. {\em J. Phys. Soc. Japan} {\bf 2012}, {\em 81}, 011009.
% Reference 10
\bibitem[Fu(2008)]{ref-journal10}
Fu, L.; Kane, C.L. Superconducting Proximity Effect and Majorana Fermions at the Surface of a Topological Insulator. {\em Phys. Rev. Lett.} {\bf 2008}, {\em 100}, 096407.
% Reference 11
\bibitem[Sau(2010)]{ref-journal11}
Sau, J.D.; Lutchyn, R.M.; Tewari, S.; Sarma, S.D. Superconducting Proximity Effect and Majorana Fermions at the Surface of a Topological Insulator. {\em Phys. Rev. Lett.} {\bf 2010}, {\em 104}, 040502.
% Reference 12
\bibitem[Sato(2009)]{ref-journal12}
Sato, M.; Fujimioto, S. Topological phases of noncentrosymmetric superconductors: Edge states, Majorana fermions, and non-Abelian statistics. {\em Phys. Rev. B.} {\bf 2009}, {\em 79}, 094504.
% Reference 13
\bibitem[Pientka(2017)]{ref-journal13}
Pientka, F.; Keselman, A.; Berg, E.; Yacoby, A.; Stern, A.; Halperin, B.I. Topological Superconductivity in a Planar Josephson Junction. {\em Phys. Rev. X} {\bf 2017}, {\em 7}, 021032.
% Reference 14
\bibitem[Chung(2011)]{ref-journal14}
Chung, S.B.; Zhang, H.J.; Qi, X.L.; Zhang, S.C. Topological superconducting phase and Majorana fermions in half-metal/superconductor heterostructures. {\em Phys. Rev. B.} {\bf 2011}, {\em 84}, 060510.
% Reference 15
\bibitem[Takei(2013)]{ref-journal15}
Takei, S.; Fregoso, B.M.; Galitski, V.; Sarma, S.D. Topological superconductivity and Majorana fermions in hybrid structures involving cuprate high-Tc superconductors. {\em Phys. Rev. B.} {\bf 2013}, {\em 87}, 014504.
% Reference 16
\bibitem[Takei(2012)]{ref-journal16}
Takei, S.; Galitski, V. Microscopic theory for a ferromagnetic nanowire/superconductor heterostructure: Transport, fluctuations, and topological superconductivity. {\em Phys. Rev. B}  {\bf 2012}, {\em 86}, 054521.
% Reference 17
\bibitem[Sau(2015)]{ref-journal17}
Sau, J.D.; Brydon, P.M.R. Bound States of a Ferromagnetic Wire in a Superconductor. {\em Phys. Rev. Lett.}  {\bf 2015}, {\em 115}, 127003.
% Reference 18
\bibitem[Dumitrescu(2015)]{ref-journal18}
Dumitrescu, E.; Roberts, B.; Tewari, S.; Sau, J.D.; Sarma, S.D.
Majorana fermions in chiral topological ferromagnetic nanowires.
{\em Phys. Rev. B}  {\bf 2015}, {\em 91}, 094505.
% Reference 19
\bibitem[Brydon(2012)]{ref-journal19}
Brydon, P.M.R.; Sarma, S.D.; Hui, H.Y.; Sau, J.D.
Topological Yu--Shiba--Rusinov chain from spin-orbit coupling.
{\em Phys. Rev. B}  {\bf 2012}, {\em 91}, 064505. 
% Reference 20
\bibitem[Perge(2013)]{ref-journal20} 
Nadj-Perge, S.; Drozdov, I.K.; Bernevig, B.A.; Yazdani, A. Proposal for realizing Majorana fermions in chains of magnetic atoms on a superconductor. {\em Phys. Rev. B} {\bf 2013}, {\em 88}, 020407.
% Reference 21
\bibitem[Jian(2014)]{ref-journal21}
Li, J.; Chen, H.; Drozdov, I.K.; Yazdani, A.; Bernevig, A.; MacDonald, A.H. Topological superconductivity induced by ferromagnetic metal chains. {\em Phys. Rev. B}  {\bf 2014}, {\em 90}, 235433.
% Reference 22
\bibitem[Rontynen(2016)]{ref-journal22}
Rontynen, J.; Ojanen, T.
Chern mosaic: Topology of chiral superconductivity on ferromagnetic adatom lattices.
{\em Phys. Rev. B}  {\bf 2016}, {\em 93}, 094521.
% Reference 23
\bibitem[Rontynen(2016)]{ref-journal23}
Poyhonen, K.; Weststrom, A.; Ojanen, T.
Topological superconductivity in ferromagnetic atom chains beyond the deep-impurity regime.
{\em Phys. Rev. B}  {\bf 2016}, {\em 93}, 014517.
% Reference 24
\bibitem[Rontynen(2016)]{ref-journal24}
Čadež, T.; Sacramento, P.D.
Zero energy modes in a superconductor
with ferromagnetic adatom chains
and quantum phase transitions.
{\em J. Phys. Condens. Matter}  {\bf 2016}, {\em 28}, 495703.
% Reference 25
\bibitem[Sticlet(2019)]{ref-journal25}
Sticlet, D.; Morari, C.
Topological superconductivity from magnetic impurities on monolayer NbSe2.
{\em Phys. Rev. B}  {\bf 2019}, {\em 100}, 075420. 
% Reference 26
\bibitem[Schneider(2021)]{ref-journal26}
Schneider, L.; Beck, P.; Posske, T.; Crawford, D.; Mascot, E.; Rachel, S.;
Wiesendanger, R.; Wiebe, J.
Topological Shiba bands in artificial spin chains on
superconductors.
{\em Nat. Phys.}  {\bf 2021}, {\em 17}, 943--948.
% Reference 27
%\bibitem[Costa(2018)]{ref-journal27}
%Costa, A.; Fabian, J.; Kochan, D.
%Connection between zero-energy Yu--Shiba--Rusinov states and $0-\pi$ %transitions in magnetic Josephson junctions.
%{\em Phys. Rev. B} {\bf 2018}, {\em 98}, 134511.
 % Reference 38
\bibitem[Livanas(2019)]{ref-journal27} 
Livanas, G.; Sigrist, M.; Varelogiannis, G.
Alternative paths to realize Majorana Fermions in Superconductor-Ferromagnet Heterostructures, {\em Sci. Rep.} {\bf 2019}, {\em 9}, 6259.
% Reference 27
\bibitem[Tokuyasu(1988)]{ref-journal28} 
Tokuyasu, T.; Sauls, J.A.; Rainer, D. Proximity effect of a ferromagnetic insulator in contact with a superconductor, {\em Phys. Rev. B}
{\bf 1988}, {\em 38}, 8823.
% Reference 28
\bibitem[Hijano(2021)]{ref-journal29} 
Hijano, A.; Ilić, S.; Rouco, M.; González-Orellana, C.; Ilyn, M.; Rogero, C.; Virtanen, P.; Heikkilä, T. T.; Khorshidian, S.; \mbox{Spies, M.; et al.}  Coexistence of superconductivity and spin-splitting fields in superconductor/ferromagnetic insulator bilayers of arbitrary thickness. {\em Phys. Rev. Res.} {\bf 2021}, {\em 3}, 023131.
% Reference 29
\bibitem[Tedrow(1986)]{ref-journal30}
Tedrow, P.M.; Tkaczyk, J.E.; Kumar, A.
Spin-Polarized Electron Tunneling Study of an Artificially Layered Superconductor with Internal Magnetic Field: EuO-Al. {\em Phys. Rev. Lett.} {\bf 1986}, {\em 56}, 1746.
% Reference 30
\bibitem[Hao(1990)]{ref-journal31}
Hao, X.; Moodera, J.S.; Meservey, R. Spin-filter effect of ferromagnetic europium sulfide tunnel barriers. {\em Phys. Rev. B} {\bf 1990}, {\em 42}, 8235.
% Reference 31
\bibitem[Moodera(2017)]{ref-journal32}
Strambini, E.; Golovach, V.N.; ; Simoni, G.D.; ; Moodera, J.S.; Bergeret, F.S.; Giazotto, F. Revealing the magnetic proximity effect in EuS/Al bilayers through superconducting tunneling spectroscopy. {\em Phys. Rev. Mater.} {\bf 2017}, {\em 1}, 054402.
% Reference 32
\bibitem[Liu(2019)]{ref-journal33}
Liu, Y.; Vaitiekėnas, S.; Martí-Sánchez, S.; Koch, C.; Hart, S.; Cui, Z.; Kanne, T.; Khan, S.A.; Tanta, R.; Upadhyay, S.; et al. Semiconductor–Ferromagnetic Insulator–Superconductor Nanowires: Stray Field and Exchange Field. {\em Nano Lett.} {\bf 2020}, {\em 20}, 456–462.
% Reference 33
\bibitem[Giroud(1998)]{ref-journal34} 
Giroud, M.; Courtois, H.; Hasselbach, K.; Mailly, D.; Pannetier, B. Superconducting proximity effect in a mesoscopic ferromagnetic wire. {\em Phys. Rev. B} {\bf 1998}, {\em 58}, R11872(R).
% Reference 34
\bibitem[Wang(2010)]{ref-journal35} 
Wang, J.; Singh, M.; Tian, M.; Kumar, N.; Liu, B.; Shi, C.; Jain, J.K.; Samarth, N.; Mallouk, T.; Chan, M.H.W.  Interplay between superconductivity and ferromagnetism in crystalline nanowires. {\em Nat. Phys.} {\bf 2010}, {\em 6}, 389–394.
% Reference 35
\bibitem[Yazdani(2019)]{ref-journal36}   
Jäck, B.; Xie, Y.; Li, J.; Jeon, S.; Bernevig, B.A.; Yazdani, A. Observation of a Majorana zero mode in a topologically protected edge channel. {\em Science}  {\bf 2019}, {\em 364}, 6447.
% Reference 36
\bibitem[Ueda(1991)]{ref-journal37}
Sigrist, M.; Ueda, K.
Phenomenological theory of unconventional superconductivity. {\em Rev. Mod. Phys.} {\bf 1991}, {\em 63}, 239.
% Reference 37
\bibitem[Tewari(2013)]{ref-journal38}
Dumitrescu, E.; Tewari, S.
Topological properties of the time-reversal-symmetric Kitaev chain and applications to organic superconductors.
{\em Phys. Rev. B} {\bf 2013}, {\em 88}, 220505(R).
% Reference 36
\bibitem[Schnyder(2008)]{ref-journal39}
Schnyder, A.P.; Ryu, S.; Furusaki, A.; Ludwig, A.W.W. Classification of topological insulators and superconductors in three spatial dimensions. {\em Phys. Rev. B} {\bf 2008}, {\em 78}, 195125.
% Reference 37
\bibitem[Tewari(2012)]{ref-journal40}
 Tewari S.; Sau J.D. Topological Invariants for Spin-Orbit Coupled Superconductor Nanowires, {\em Phys. Rev. Lett.} {\bf 2012}, {\em 109}, 150408.
% Reference 39
\bibitem[Peng(2015)]{ref-journal41}
Peng, Y.; Pientka, F.; Glazman, L.I.; von Oppen, F.
Strong Localization of Majorana End States in Chains of Magnetic Adatoms.
{\em Phys. Rev. Lett.} {\bf 2015}, {\em 114}, 106801.
% Reference 40
\bibitem[Law(2009)]{ref-journal42}
Law, K.T.; Lee, P.A.; Ng, T.K. Majorana Fermion Induced Resonant Andreev Reflection. {\em Phys. Rev. Lett.} {\bf 2009}, {\em 103}, 237001.
% Reference 40
{\bibitem[Law(2009)]{ref-journal43}
He, J.J.; Ng, T.K.; Lee, P.A.; Law, K.T. Selective Equal-Spin Andreev Reflections Induced by Majorana Fermions.
{\em Phys. Rev. Lett.} {\bf 2014}, {\em 112}, 037001.}
% Reference 41
\bibitem[Mourik(2012)]{ref-journal44} 
Mourik, V.; Zuo, K.; Frolov, S.M.; Plissard, S.R.; Bakkers, E.P.A.M.; Kouwenhoven, L.P.  Signatures of Majorana Fermions in Hybrid Superconductor-Semiconductor Nanowire Devices. {\em Science} {\bf 2012}, {\em 336}, 1003.
% Reference 42
\bibitem[Sun(2016)]{ref-journal45} 
Sun, H.H.; Zhang, K.W.; Hu, L.H.; Li, C.; Wang, G.Y.; Ma, H.Y.; Xu, Z.A.; Gao, C.L.; Guan, D.D.; Li, Y.Y.; et al. Majorana Zero Mode Detected with Spin Selective Andreev Reflection in the Vortexof a Topological Superconductor. {\em Phys. Rev. Lett.} {\bf 2016}, {\em 116}, 257003.
% Reference 45
{\bibitem[Zhang(2016)]{ref-journal46} 
Hu, L.H.; Li, C.; Xu, D.H.; Zhou, Y.; Zhang, F.Z. Theory of spin-selective Andreev reflection in the vortex core of a topological superconductor.
{\em Phys. Rev. B} {\bf 2016}, {\em 94}, 224501.}
% Reference 43
\bibitem[Perge(2014)]{ref-journal47} 
Nadj-Perge, S.; Drozdov, I.K.; Li, J.; Chen, H.; Jeon, S.; Seo, J.; MacDonald, A.H.; Bernevig, B.A.; Yazdani, A. Observation of Majorana fermions in ferromagnetic atomic chains on a superconductor. {\em Science} {\bf 2014}  {\em 346}, 602.
% Reference 44
\bibitem[Wei(2019)]{ref-journal48} 
Wei, P.; Manna, S.; Eich, M.; Lee, P.; Moodera, J. Superconductivity in the Surface State of Noble Metal Gold and its Fermi Level Tuning by EuS Dielectric. {\em Phys. Rev. Lett.} {\bf 2019}, {\em 122}, 247002.
% Reference 45
\bibitem[He(2017)]{ref-journal49} 
He, Q.L.; Pan, L.; Stern, A.L.; Burks, E.C.; Che, X.; Yin, G.; Wang, J.; Lian, B.; Zhou, Q.; Choi, E.S.; et al. Chiral Majorana fermion modes in a quantum anomalous Hall insulator–superconductor structure. {\em Science} {\bf 2017}, {\em 357}, 294.
% Reference 46
\bibitem[Wang(2018)]{ref-journal50} 
Wang, D.; Kong, L.; Fan, P.; Chen, H.; Zhu, S.; Liu, W.; Cao, L.; Sun, Y.; Du, S.; Schneeloch, J.; et al. Evidence for Majorana bound states in an iron-based superconductor. {\em Science} {\bf 2018}, {\em 362}, 333.
% Reference 47
\bibitem[Fornieri(2019)]{ref-journal51}
 Fornieri, A.; Whiticar, A.; Setiawan, F.; Portolés, E.; Drachmann, A.; Keselman, A.; Gronin, S.; Thomas, C.; Wang, T.; \mbox{Kallaher, R.; et al.}  Evidence of topological superconductivity in planar Josephson junctions. {\em Nature} {\bf 2019}, {\em 569}, 89–92.
% Reference 48
\bibitem[Brouwer(2011)]{ref-journal52}
Brouwer, P.W.; Duckheim, M.; Romito, A.; von Oppen, F.
Probability Distribution of Majorana End-State Energies in Disordered Wires,
{\em Phys. Rev. Lett.} {\bf 2011}, {\em 107}, 196804.
% Reference 49\
\bibitem[Yu(1969)]{ref-journal53}
Yu, L. Bound state in superconductors with paramagnetic impurities. %MDPI: It is not permitted if there are three references, please divide them here, also in the main text
 {\em Acta Phys. Sin.} {\bf 1965}, {\em 21}, 75.
\bibitem[Shiba(1969)]{ref-journal54}
Shiba, H. Classical spins in superconductors. {\em Prog. Theor. Phys.} {\bf 1968}, {\em 40}, 435.
\bibitem[Rusinov(1969)]{ref-journal55}
Rusinov, A.I. Superconductivity near a paramagnetic impurity. {\em Zh. Eksp. Teor. Fiz. Pisma. Red.} {\bf 1969}, {\em 9}, 146.
% Reference 50
%\bibitem[Dumitrescu(2015)]{ref-journal53}
%Dumitrescu, E.; Roberts, B.; Tewari, s. Sau, J.S.; and Das Sarma S.
%Majorana fermions in chiral topological ferromagnetic nanowires,
%{\em Phys. Rev. B} {\bf 2015}, {\em 91}, 094505.
% Reference 36
%\bibitem[Avci(2017)]{ref-journal36}
%Avci, C., Quindeau, A., Pai, CF., et al. Current-induced switching in a magnetic insulator, {\em Nature Mater} {\bf 2017}, {\em 16}, 309.
 % Reference 37
%\bibitem[Chiba(2008)]{ref-journal37}
%D. Chiba, M. Sawicki, Y. Nishitani, et al. \href{https://doi.org/10.1038/nature07318}{Nature 455, 515 (2008).}
% Reference 38
%\bibitem[Chiba2(2010)]{ref-journal38} 
%D. Chiba, Y. Nakatani, F. Matsukura, and H. Ohno
%\href{http://dx.doi.org/10.1063/1.3428959}{Appl. Phys. Lett. 96, 192506 (2010)}.
%\bibitem[Duckheim(2011)]{ref-journal39}
%Duckheim, M., and Brouwer, P. W.,
%Andreev reflection from noncentrosymmetric superconductors and Majorana bound-state generation in half-metallic ferromagnets.
%{\em Phys. Rev. B} {\bf 2011}, {\em 83}, 054513.




%\bibitem[Law(2009)]{ref-journal16}
%K. T. Law, Patrick A. Lee, and T. K. Ng, Majorana Fermion Induced Resonant Andreev Reflection. {\em Phys. Rev. Lett.} {\bf 2009}, {\em 103}, 237001.
%\bibitem[Kwon(2004)]{ref-journal17}
%Kwon, HJ., Sengupta, K. and Yakovenko, V.M., Fractional ac Josephson effect in p- and d-wave superconductors. {\em Eur. Phys. J. B} {\bf 2004}, {\em 37}, 349-361.
%\bibitem[Jiang(2013)]{ref-journal18}
%Liang Jiang et al., Magneto-Josephson effects in junctions with Majorana bound states. {\em Phys. Rev. B} {\bf 2013}, {\em 87}, 075438.
%\bibitem[Mourik(2012)]{ref-journal19}
%V. Mourik et al., Signatures of Majorana Fermions in Hybrid Superconductor-Semiconductor Nanowire Devices. {\em Science} {\bf 2012}, {\em 336}, 1003.
%\bibitem[Perge(2014)]{ref-journal20}
%Stevan Nadj-Perge et al., Observation of Majorana fermions in ferromagnetic atomic chains on a superconductor. {\em Science} {\bf 2014}, {\em 346}, 602.
%\bibitem[Manna(2014)]{ref-journal21}
%Sujit Manna, et al., Signature of a pair of MZMs in superconducting gold surface states. {\em PNAS} {\bf 2020}, {\em 117}, 8775-8782.
%\bibitem[Rokhinson(2012)]{ref-journal22}
%Rokhinson, L. P., Liu, X. and Furdyna, J. K., The fractional a.c. Josephson effect in a semiconductor-superconductor nanowire as a signature of Majorana particles. {\em Nature Phys} {\bf 2012}, {\em 8}, 795-799.
%\bibitem[Wiedenmann(2016)]{ref-journal23}
%Wiedenmann, J. et al., $4\pi$-periodic Josephson supercurrent in HgTe-based topological Josephson junctions. {\em Nat Commun} {\bf 2016}, {\em 7}, 10303.
%\bibitem[Fornieri(2019)]{ref-journal24}
%A. Fornieri et al., Evidence of topological superconductivity in planar Josephson junctions. {\em Nature} {\bf 2019}, {\em 569}, 89-92.
%\bibitem[Pan(2020)]{ref-journal25}
%Haining Pan and S. Das Sarma, Physical mechanisms for zero-bias conductance peaks in Majorana nanowires. {\em Phys. Rev. Research} {\bf 2020}, {\em 2}, 013377.
%\bibitem[Pan(2019)]{ref-journal26}
%Ching-Kai Chiu and S. Das Sarma, Fractional Josephson effect with and without Majorana zero modes. {\em Phys. Rev. B} {\bf 2019}, {\em 99}, 035312.
%\bibitem[Pan(2021)]{ref-journal27}
%Dartiailh, M.C., Cuozzo, J.J., Elfeky, B.H. et al., Missing Shapiro steps in topologically trivial Josephson junction on InAs quantum well. {\em Nat Commun} {\bf 2021}, {\em 12}, 78.
%\bibitem[Beenaker(2021)]{ref-journal28}
%C. W. J. Beenakker, P. Baireuther, Y. Herasymenko, I. Adagideli, L. Wang and A. R. Akhmerov, Deterministic creation and braiding of chiral edge vortices. {\em Phys. Rev. Lett.} {\bf 2019}, {\em 122}, 146803.
%\bibitem[Alicea(2011)]{ref-journal29}
%J. Alicea, Y. Oreg, G. Refael, F. von Oppen, and M. P. A. Fisher, Non-Abelian statistics and topological quantum information processing in 1D wire networks. {\em Nature Phys.} {\bf 2011}, {\em 7}, 412-417.
%\bibitem[Harper(2019)]{ref-journal30}
%Fenner Harper, Aakash Pushp, and Rahul Roy, Majorana braiding in realistic nanowire Y-junctions and tuning forks. {\em Phys. Rev. Research} {\bf 2019}, {\em 1}, 033207.
%\bibitem[Tutschku(2020)]{ref-journal31}
%C. Tutschku, R. W. Reinthaler, C. Lei, A. H. MacDonald, and E. M. Hankiewicz, Majorana-based quantum computing in nanowire devices. {\em Phys. Rev. B} {\bf 2020}, {\em 102}, 125407.
%\bibitem[Malciu(2018)]{ref-journal32}
%Corneliu Malciu, Leonardo Mazza, and Christophe Mora, Braiding Majorana zero modes using quantum dots. {\em Phys. Rev. B} {\bf 2018}, {\em 98}, 165426.
%\bibitem[Heck(2012)]{ref-journal33}
%B van Heck et al., Coulomb-assisted braiding of Majorana fermions in a Josephson junction array. {\em New J. Phys.} {\bf 2012}, {\em 14}, 035019.
%\bibitem[Kim(2021)]{ref-journal34}
%M. D. Kim, Circulator function in a Josephson junction circuit and braiding of Majorana zero modes. {\em Sci Rep} {\bf 2021}, {\em 11}, 1826.
%\bibitem[Li(2016)]{ref-journal35}
%Li, J., Neupert, T., Bernevig, B. et al., Manipulating Majorana zero modes on atomic rings with an external magnetic field. {\em Nat Commun} {\bf 2016}, {\em 7}, 10395.
%\bibitem[Hedge(2020)]{ref-journal36}
%Suraj S. Hegde et al., A topological Josephson junction platform for creating, manipulating, and braiding Majorana bound states. {\em Annals of Physics} {\bf 2020}, {\em 423}, 168326.
%\bibitem[Kotetes(2013)]{ref-journal37}
%P. Kotetes, G. Schön and A. Shnirman, Engineering and manipulating topological qubits in 1D quantum wires. {\em Journal of the Korean Physical Society} {\bf 2013}, {\em 62}, 1558-1563.
%\bibitem[Cuoco(2017)]{ref-journal38}
%P. Marra and M. Cuoco, Controlling Majorana states in topologically inhomogeneous superconductors. {\em Phys. Rev. B} {\bf 2017}, {\em 95}, 140504.
%\bibitem[Kim(2015)]{ref-journal39}
%Se Kwon Kim, Control and braiding of Majorana fermions bound to magnetic domain walls. {\em Phys. Rev. B} {\bf 2015}, {\em 92}, 020412.



\end{thebibliography}
\end{document}